\documentclass[aps,pre,twocolumn,bibnotes,groupedaddress]{revtex4-1}

\usepackage{amsmath,amsthm,amsfonts,amssymb,bm}
\usepackage{graphicx,subfigure}
\usepackage{color}
\usepackage{natbib}
\usepackage{mciteplus}

\newcommand{\eigen}{\lambda}
\newcommand{\base}[1]{\hat{{#1}}}
\newcommand{\gdot}{\dot{\gamma}}
\newcommand{\gdotbar}{\overline{\dot{\gamma}}}

\newcommand{\beqn}{\begin{equation}}
\newcommand{\eeqn}{\end{equation}}
\newcommand{\beqna}{\begin{eqnarray}}
\newcommand{\eeqna}{\end{eqnarray}}

\newcommand{\taur}{\tau_R}
\newcommand{\taud}{\tau_d}

\newcommand{\vecv}[1]{\bm{{#1}}}
\newcommand{\tens}[1]{\bm{{#1}}}

\newcommand{\visc}{\tens{W}} 

\newcommand{\wxy}{W_{xy}}
\newcommand{\wyy}{W_{yy}}

\usepackage{graphicx,amsmath,color,amsfonts}

\begin{document}
 
\title{Shear banding in large amplitude oscillatory shear (LAOStrain
  and LAOStress) of polymers and wormlike micelles}

\author{K. A. Carter} 
\author{J. M. Girkin} 
\author{S. M. Fielding} 

\affiliation{Department of Physics, Durham University, Science
  Laboratories, South Road, Durham, DH1 3LE, UK}
  
\date{\today}
\begin{abstract} { We investigate theoretically shear banding in large
    amplitude oscillatory shear (LAOS) of polymeric and wormlike
    micellar surfactant fluids. In LAOStrain we observe banding at low
    frequencies and sufficiently high strain rate amplitudes in fluids
    for which the underlying stationary constitutive curve of shear
    stress as a function of shear rate is non-monotonic. This is the
    direct (and relatively trivial) analogue of quasi steady state
    banding seen in slow strain rate sweeps along the flow curve. At
    higher frequencies and sufficiently high strain amplitudes we
    report a different but related phenomenon, which we call `elastic'
    shear banding. This is associated with an overshoot in the elastic
    (Lissajous-Bowditch) curve of stress as a function of strain and
    we suggest that it might arise rather widely even in fluids that
    have a monotonic underlying constitutive curve, and so do not show
    steady state banding if under a steadily applied shear flow. It is analogous to the elastic banding
    triggered by stress overshoot in a fast shear startup predicted
    previously in~\cite{Moorcroft2013}, but could be more
    readily observable experimentally in this oscillatory protocol due
    to its recurrence in each half cycle.  In LAOStress we report
    shear banding in fluids that shear thin strongly enough to have
    either a negatively, or weakly positively, sloping region in the
    underlying constitutive curve, noting again that fluids in the
    latter category do not display steady state banding in a steadily applied flow. This banding
    is triggered in each half cycle as the stress magnitude transits the region
    of weak slope in a upward direction, such that the fluid
    effectively yields. It is strongly reminiscent of the transient
    banding predicted previously in step
    stress~\cite{Moorcroft2013}.  Our numerical calculations are
    performed in the Rolie-poly model of polymers and wormlike
    micelles, but we also provide arguments suggesting that our
    results should apply more widely.  Besides banding in the shear
    strain rate profile, which can be measured by velocimetry, we also
    predict banding in the shear and normal stress components,
    measurable by birefringence. As a backdrop to understanding the
    new results on shear banding in LAOS, we also briefly review
    earlier work on banding in other time-dependent protocols,
    focusing in particular on shear startup and step stress.}
\end{abstract}
\maketitle

\newpage

\section{Introduction}

Many complex fluids display shear banding, in which a state of
initially homogeneous shear flow gives way to the formation of
coexisting bands of differing shear rate, with layer normals in the
flow-gradient direction. For recent reviews,
see~\cite{Olmsted2008,Manneville2008a,Fielding2014,Divoux2015}.
Following its early observation in wormlike micellar surfactant
solutions~\cite{Britton1997}, over the past two decades shear banding
has been seen in virtually all the major classes of complex fluids and
soft solids. Examples include microgels~\cite{Divoux2010},
clays~\cite{Martin2012}, emulsions~\cite{Coussot2002}
foams~\cite{Rodts2005}, lamellar surfactant
phases~\cite{Salmonetal2003a}, triblock
copolymers~\cite{Berretetal2001a,Mannevilleetal2007a}, star
polymers~\cite{Rogers2008}, and -- subject to ongoing
controversy~\cite{Wangetal2003a,Wangetal2006c,Wangetal2008a,Li2013,Wang2014,Li2015,Wang2011,Wangetal2008a} -- 
linear polymers.

Prior to about 2010, the majority of studies of shear banding focused
on conditions of a steadily applied shear flow. The criterion for the
presence of steady state banding in this case is well known: that the
underlying homogeneous constitutive curve of shear stress as a
function of shear rate has a regime of negative slope. (In some cases
of strong concentration coupling shear banding can arise even for a
monotonic constitutive curve~\cite{Fielding2003a}, but we
do not consider that case here.)  Such a regime is predicted by the
original tube theory of Doi and Edwards for non-breakable
polymers~\cite{DoiEdwards}, and by the reptation-reaction model of
wormlike micellar surfactants~\cite{Cates1990}.  It is
straightforward to show that a state of initially homogeneous shear
flow is linearly unstable, in this regime of negative constitutive
slope, to the formation of shear bands~\cite{Yerushalmi1970}.  The
composite steady state flow curve of shear stress as a function of
shear rate then displays a characteristically flat plateau regime, in
which shear bands are observed.

From an experimental viewpoint, the evidence for steady state shear
banding under a steadily applied shear flow is now overwhelming in the case
of wormlike micelles. For reviews,
see~\cite{Cates2007,Berret2005}. For linear unbreakable
polymers the issue remains controversial, as recently reviewed in
Ref.~\cite{Snijkers2015}. In particular the original
Doi-Edwards model did not account for a process known as convective
constraint release
(CCR)~\cite{Marrucci1996,Ianniruberto2014a,Ianniruberto2014}.
Since CCR (which we describe below) was proposed, there has been an
ongoing debate about its efficacy in potentially eliminating the
regime of negative constitutive slope and restoring a monotonic
constitutive curve, thereby eliminating steady state banding.
However, a non-monotonic constitutive curve and associated steady
state shear banding has been seen in molecular dynamics simulations of
polymers~\cite{Likhtmanetal2012}, for long enough chain lengths.  It
is important to note, though, that the polydispersity that is often
present in practice in unbreakable polymers also tends to restore
monotonicity.

Besides the conditions of steady state flow just described, many flows
of practical importance involve a strong time dependence.  In view of
this, a natural question to ask is whether shear banding might also
arise in these time-dependent flows and, if so, under what conditions.
Over the past decade, a body of experimental data has accumulated to
indicate that it does indeed occur: in shear
startup~\cite{Divoux2010, Divoux2011a, Wangetal2009a,
  Huetal2007a, Wangetal2008a, Martin2012}, following a step
strain (in practice a rapid strain ramp)~\cite{Wangetal2010a,
  Boukany2009a, Wangetal2006a, Fang2011, Wangetal2007a,
  Archer1995, Wangetal2008c}, and following a step
stress~\cite{Gibaud2010,Divoux2011,
  Wangetal2009a,Huetal2007a, Wangetal2003a, Hu2008,
  Wangetal2008c, Hu2005, Hu2010}.

Consistent with this growing body of experimental evidence,
theoretical
considerations~\cite{Moorcroft2013,Moorcroft2014,Moorcroft2011,Adamsetal2011,Manningetal2009a}
also suggest that shear banding might arise rather generically in
flows with a sufficiently strong time-dependence, even in fluids that
have a monotonically increasing constitutive curve and so do not
display steady state banding under conditions of a continuously
applied shear.  Indeed, the calculations to date suggest that the set
of fluids that show banding in steady state is only a subset of those
that exhibit banding in time-dependent flows. In view of this,
although the question concerning the existence or otherwise of steady
state shear banding in polymers remains an important one, the
resolution of that controversy is likely to be of less practical
importance to the broader issue of whether shear banding arises more
generally in time-dependent flows.

In the last five years progress has been made in establishing
theoretically, separately for each of the time-dependent flow
protocols listed above (shear startup, step strain and step stress), a
fluid-universal criterion~\cite{Moorcroft2013} for the onset of
shear banding, based on the shape of the time-dependent rheological
response function for the particular protocol in question. We now
briefly review these criteria as backdrop to understanding the results
that follow below for shear banding in large amplitude oscillatory
shear (LAOS).

In shear startup (the switch-on at some time $t=0$ of a constant shear
rate $\gdot$), the onset of banding is closely associated with the
presence of an
overshoot~\cite{Moorcroft2013,Moorcroft2014,Moorcroft2011,Adamsetal2011,Manningetal2009a}
in the startup signal of stress as a function of time (or equivalently
as a function of strain), as it evolves towards its eventual steady
state on the material's flow curve. This concept builds on the early
insight of Ref.~\cite{Marrucci1983}. The resulting bands may, or may
not, then persist to steady state, according to whether or not the
underlying constitutive curve of stress as a function of strain rate
is non-monotonic. This tendency of a startup overshoot to trigger
banding was predicted on the basis of fluid-universal analytical
calculations in Ref.~\cite{Moorcroft2013}, and has been
confirmed in numerical simulations of polymeric fluids (polymer solutions, polymer melts and
wormlike micelles)~\cite{Moorcroft2014,Adamsetal2011},
polymer glasses~\cite{Fielding2013} and soft glassy materials
(dense emulsions, microgels, foams, {\it
  etc.})~\cite{Moorcroft2011,Manningetal2009a,Jagla2010}.
It is consistent with experimental observations in wormlike micellar
surfactants~\cite{Hu2008, Wangetal2008c},
polymers~\cite{Wangetal2008a, Wangetal2008b, Wangetal2009a,
  Wangetal2009b, Boukany2009a, Wangetal2003a, Huetal2007a,
  Wangetal2009d}, carbopol
gels~\cite{Divoux2011a,Divoux2010} and Laponite
clay suspensions~\cite{Martin2012}.

Following the imposition of a step stress in a previously undeformed
sample, the onset of shear banding is closely associated with the
existence of a regime of simultaneous upward slope and upward
curvature in the time-differentiated creep response curve of shear
rate as a function of
time~\cite{Moorcroft2013,Fielding2014}.  This criterion
was also predicted on the basis of fluid-universal analytical
calculations in Ref.~\cite{Moorcroft2013}, and has been
confirmed in numerical simulations of polymeric
fluids~\cite{Moorcroft2014} and soft glassy
materials~\cite{Fielding2014}.  It is consistent with
experimental observations in polymers~\cite{Wangetal2009a,Huetal2007a,Wangetal2003a, Hu2008,
  Wangetal2008c, Hu2005, Hu2010, Wangetal2009d}, carbopol
microgels~\cite{Divoux2011} and carbon black
suspensions~\cite{Gibaud2010}.

In the shear startup and step stress experiments just described, the
time-dependence is inherently transient in nature: after (typically)
several strain units, the system evolves to its eventual steady state
on the material's flow curve.  In any such protocol, for a fluid with
a monotonic constitutive curve that precludes steady state banding,
any observation of banding is predicted to be limited to this regime
of time-dependence following the inception of the flow.  That poses an
obvious technical challenge to experimentalists: of imaging the flow
with sufficient time-resolution to detect these transient bands.  This
is particularly true for a polymeric fluid with a relatively fast
relaxation spectrum.  For soft glassy materials, in contrast, the
dynamics are typically much slower and any bands associated with the
onset of flow, though technically transient, may persist for a
sufficiently long time to be mistaken for the material's ultimate
steady state response for any practical
purpose~\cite{Moorcroft2011,Fielding2014}.

In the past decade, the rheological community has devoted considerable
attention to the of study large amplitude oscillatory shear (LAOS).
For a recent review, see Ref.~\cite{Hyun2011}. In this
protocol, the applied flow has the form of a sustained oscillation and
is therefore perpetually time-dependent, in contrast to the transient
time-dependence of the shear startup and step stress protocols just
described.  But by analogy with the predictions of transient shear
banding in shear startup and step stress, a sustained oscillatory flow
might (in certain regimes that we shall discuss) be expected to
repeatedly show banding at certain phases of the cycle, or even to
show sustained banding round the whole cycle.  Importantly, again by
analogy with our knowledge of shear startup and step stress, this
effect need not be limited to fluids with a non-monotonic constitutive
curve that show steady state banding in a continuously applied shear
flow, but might instead arise as a natural consequence of the
time-dependence inherent to the oscillation.

Indeed, a particularly attractive feature of LAOS is that the severity
of the flow's time-dependence, relative to the fluid's intrinsic
characteristic relaxation timescale $\tau$, can be tuned by varying
the frequency $\omega$ of the applied oscillation. A series of LAOS
experiments can thereby explore the full range between steady state
behaviour in the limit $\omega\to 0$, where the oscillation
effectively corresponds to a repeated series of quasi-static sweeps up
and down the flow curve, and strongly time-dependent behaviour for
$\omega > 1/\tau$. A fluid with a non-monotonic underlying
constitutive curve that admits steady state banding is then clearly
expected to exhibit banding in the limit of $\omega\to 0$, as the
shear rate quasi-statically transits the plateau in the steady state
flow curve. In contrast, a monotonic constitutive curve precludes
banding for $\omega\to 0$.  Crucially, though, as noted above, the
absence of banding in steady state conditions does not rule out the
possibility of banding in flows with a strong enough time-dependence,
$\omega \gtrapprox O(1/\tau)$.

Indeed, intuitively, a square-wave caricature of a large amplitude
oscillatory shear strain (LAOStrain) experiment points to a perpetual
switching between a shear startup like process in the forward
direction, followed by `reverse startup' in the opposite direction.
Any regime in which these startup-like events are associated with an
overshoot in the associated curve of stress as a function of strain
then strongly suggests the possibility of shear banding during those
quasi-startup parts of the cycle, by analogy with the criterion for
banding in a true shear startup from rest.  In the same spirit, a
square-wave caricature of a large amplitude oscillatory shear stress
(LAOStress) experiment indicates a perpetually repeated series of step
stress events, jumping between positive and negative stress values,
and so admitting the possibility of shear banding if the criterion for
banding following a step stress is met.

In practice, of course, LAOS is more complicated than the caricatures
just described and the criteria for banding in shear startup and step
stress might only be expected to apply in certain limiting regimes.
Nonetheless, in what follows we shall show that many of our results
for banding in LAOStrain and LAOStress can, to a large extent, be
understood within the framework of these existing criteria for the
simpler time-dependent protocols.

Experimentally, shear banding has indeed been observed in LAOS: in
polymer solutions~\cite{Wangetal2006c}, dense
colloids~\cite{Cohen2006}, and also in wormlike micellar
surfactants that are known to shear band in steady
state~\cite{Dimitriou2012,KateGurnon2012,Gurnon2014a}.

From a theoretical viewpoint, several approaches to the interpretation
of LAOS data have been put forward in the
literature~\cite{Hyun2011}. These include Fourier transform
rheology~\cite{Wilhelm2002}; measures for quantifying
Lissajous-Bowditch curves (defined below) in their elastic
representation of stress versus strain, or viscous representation of
stress versus strain rate~\cite{Tee1975}; a decomposition
into characteristic sine, square and triangular wave prototypical
response functions~\cite{Klein2007,Klein2008};
decomposition into elastic and viscous stress contributions using
symmetry arguments~\cite{Cho2005}; Chebyshev series
expansions of these elastic and viscous
contributions~\cite{Ewoldt2008a}; and interpretations of the
LAOS cycle in terms of a sequence of physical
processes~\cite{Rogers2012,Rogers2011}.

However, many of these existing theoretical studies assume either
explicitly or implicitly that the flow remains homogeneous, and
thereby fail to take account of the possibility of shear banding. An
early exception can be found in
Refs.~\cite{Zhou2010,Zhou2008}, which studied a
model of wormlike micellar surfactants with a non-monotonic
constitutive curve in LAOStrain.  Another exception is in the paper of
Adams and Olmsted~\cite{Adams2009}, which recognised that
shear banding can arise even in the absence of any non-monotonicity in
the underlying constitutive curve.

The work that follows here builds on the remarkable insight of these
early papers, in carrying out a detailed numerical study of shear
banding in LAOStrain and LAOStress within the Rolie-poly
model~\cite{Grahametal2003a} of polymers and wormlike
micellar surfactant solutions.  Consistent with the above discussion,
in LAOStrain we observe banding at low frequencies $\omega\to 0$ and
sufficiently high strain rate amplitudes $\gdot\gtrapprox 1/\tau$ in
fluids for which the underlying constitutive curve of shear stress as
a function of shear rate is non-monotonic. At higher frequencies
$\omega=O(1/\tau)$ and for sufficiently high strain amplitudes
$\gamma\gtrapprox 1$ we instead see `elastic' shear banding associated
with an overshoot in the elastic curve of stress as a function of
strain, in close analogy with the elastic banding predicted in a fast
shear startup
experiment~\cite{Moorcroft2013,Moorcroft2014,Adamsetal2011,Adams2009}.
Importantly, we show that this elastic banding arises robustly even in
a wide range of model parameter space for which the underlying
constitutive curve is monotonic, precluding steady state banding.

In LAOStress we observe banding in fluids that shear thin sufficiently
strongly to have either a negatively, or weakly positively, sloping
region in the underlying constitutive curve. We emphasise again that
fluids in the latter category do not display steady state banding, and
therefore that, for such fluids, the banding predicted in LAOStress is
a direct result of the time-dependence of the applied flow. In this
case the banding is triggered in each half cycle as the stress magnitude
transits in an upward direction the region of weak slope and the
strain rate magnitude increases dramatically such that the material effectively
yields. This is strongly reminiscent of the transient banding
discussed previously in step
stress~\cite{Moorcroft2013,Moorcroft2014}.

While it would be interesting to interpret our findings within one (or
more) of the various mathematical methodologies for analysing LAOS
discussed above (and in particular to consider the implications of
banding for the presence of higher harmonics in the output rheological
time series), in the present manuscript we focus instead on the
physical understanding that can be gained by considering the shapes of
the signals of stress versus strain or strain rate (in LAOStrain) and
strain rate versus time (in LAOStress).  In that sense, this work is
closest in spirit to the sequence of physical processes (SPP) approach
of Refs.~\cite{Rogers2012,Rogers2011} (which did
not, however, explicitly consider heterogeneous response).  In
particular, we seek to interpret the emergence of shear banding in
LAOS on the basis of the existing criteria for the onset of banding in
the simpler time-dependent protocols of shear startup and step
stress~\cite{Moorcroft2013}.

The paper is structured as follows. In Sec.~\ref{sec:models} we
introduce the model, flow geometry and protocols to be considered.
Sec.~\ref{sec:methods} outlines the calculational methods that we
shall use.  Sec.~\ref{sec:recap} contains a summary of previously
derived linear instability criteria for shear banding in steady shear,
fast shear startup and step shear stress protocols, with the aim of providing a
backdrop to understanding shear banding in oscillatory protocols.  In
Secs.~\ref{sec:LAOStrain} and~\ref{sec:LAOStress} we present our
results for LAOStrain and LAOStress respectively, and discuss their
potential experimental verification.  Finally
Sec.~\ref{sec:conclusions} contains our conclusions and an outlook for
future work.

\section{Model, flow geometry and protocols}
\label{sec:models}

We write the stress $\tens{\Sigma}(\tens{r},t)$ at any time $t$ in a
fluid element at position $\tens{r}$ as the sum of a viscoelastic
contribution $\tens{\sigma}(\tens{r},t)$ from the polymer chains or
wormlike micelles, a Newtonian contribution characterised by a
viscosity $\eta$, and an isotropic contribution with pressure
$p(\tens{r},t)$:
\beqn
\tens{\Sigma} = \tens{\sigma} + 2 \eta \tens{D} - p\tens{I}.
\label{eqn: total_stress_tensor}
\eeqn
The Newtonian stress $2 \eta \tens{D}(\tens{r},t)$ may arise from the
presence of a true solvent, and from any polymeric
degrees of freedom considered fast enough not to be ascribed their own
viscoelastic dynamics. The symmetric strain rate tensor $\tens{D} =
\frac{1}{2}(\tens{K} + \tens{K}^T)$ where $K_{\alpha\beta} =
\partial_{\beta}v_{\alpha}$ and $\tens{v}(\tens{r},t)$ is the fluid
velocity field.  

We consider the zero Reynolds number limit of creeping flow, in which
the condition of local force balance requires the stress field
$\tens{\Sigma}(\tens{r},t)$ to be divergence free:
\beqn
\vecv{\nabla}\cdot\,\tens{\Sigma} = 0.
\label{eqn: force_balance}
\eeqn
The pressure field $p(\tens{r},t)$ is determined by enforcing that the
flow remains incompressible:
\beqn
\label{eqn: incomp}
\vecv{\nabla}\cdot\vecv{v} = 0.
\eeqn

The viscoelastic stress is then written in terms of a constant elastic
modulus $G$ and a tensor $\visc(\tens{r},t)$ characterising the
conformation of the polymer chains or wormlike micelles,
$\tens{\sigma} = G\, (\visc - \tens{I})$.  We take the dynamics of
$\visc$ to be governed by the Rolie-poly (RP)
model~\cite{Grahametal2003a} with
\begin{widetext}
\beqna
\partial_t{\visc}+\tens{v}\cdot\nabla\tens{\visc} &=& \tens{K} \cdot \visc + \visc \cdot \tens{K}^T - \frac{1}{\taud}\left(\visc - \tens{I}\right) 
- \frac{2(1-A)}{\taur}\left[\, \visc + \beta A^{-2\delta}\left(\visc - \tens{I}\right) \right] + D\nabla^2\visc,
\label{eqn: rolie-poly_tensor}
\eeqna
\end{widetext}
in which $A = \sqrt{3/T\,}$ with trace $T = \text{tr}\,\tens{\visc}$.
This RP model is a single mode simplification of the GLAMM model
\cite{GLAMM}, which provides a microscopically derived stochastic
equation for the dynamics of a test chain (or micelle) in its mean
field tube of entanglements with other chains.  The timescale $\taud$
sets the characteristic time on which a chain escapes its tube by
means of 1D curvilinear diffusion along the tube's contour, known as
reptation, allowing the molecular orientation to refresh itself. The
Rouse timescale $\taur$ sets the shorter time on which chain stretch,
as characterised by $T = \text{tr}\,\tens{\visc}$, relaxes.  The ratio
$\taud/\taur=3Z$, where $Z$ is the number of entanglements per chain.
The parameters $\beta$ and $\delta$ govern a phenomenon known as
convective constraint
release~\cite{Marrucci1996,Ianniruberto2014a,Ianniruberto2014}
(CCR), in which the relaxation of the stretch of a test chain has the
effect of also relaxing entanglement points, thereby facilitating the
relaxation of tube orientation.  The diffusive term $D\nabla^2\visc$
added to the right hand side of Eqn.~\ref{eqn: rolie-poly_tensor} is
required to account for the slightly diffuse nature of the interface
between shear bands~\cite{Luetal2000a}: without it the shear
rate would be discontinuous across the interface, which is unphysical.

Using this model we will consider shear flow between infinite flat
parallel plates at $y = \{0,L\}$, with the top plate moving in the
$\vecv{\hat{x}}$ direction at speed $\gdotbar(t) L$. We assume
translational invariance in the flow direction $\vecv{\hat{x}}$ and
vorticity direction $\vecv{\hat{z}}$ such that the fluid velocity can
be written as $\vecv{v} = v(y,t)\vecv{\hat{x}}$. The local shear rate
at any position $y$ is then given by
\beqn
\gdot(y,t) = \partial_{y}v(y,t),
\eeqn
and the spatially averaged shear rate
\beqn
\gdotbar(t) = \frac{1}{L}\int_{0}^{L} \gdot(y,t)dy.
\eeqn
Such a flow automatically satisfies the constraint of
incompressibility, Eqn.~\ref{eqn: incomp}.  The force balance
condition, Eqn.~\ref{eqn: force_balance}, further demands that the
total shear stress is uniform across the cell, in the planar flow
situation considered here, giving $\partial_{y}\Sigma_{xy} =0$. The
viscoelastic and Newtonian contributions may, however, each depend on
space provided their sum remains uniform:
\beqn
\Sigma_{xy}(t) = G\wxy(y,t) + \eta \gdot(y,t).
\label{eqn: shear_stress}
\eeqn

For such a flow, the RP model can be written componentwise as
\begin{widetext}
\beqna
\dot{W}_{xy}  &=& \gdot \wyy - \frac{\wxy}{\taud} - \frac{2(1-A)}{\taur}(1+ \beta A)\wxy + D\partial_y^2 \wxy, \nonumber\\
\dot{W}_{yy}  &=& - \frac{\wyy-1}{\taud} - \frac{2(1-A)}{\taur}\left[\wyy+ \beta A(\wyy-1)\right]+ D\partial_y^2\wyy,\nonumber\\
\dot{T}     &=&  2\dot{\gamma}\wxy  - \frac{T-3}{\taud} - \frac{2(1-A)}{\taur}\left[T + \beta A(T - 3)\right]+ D\partial_y^2 T.\quad \quad
\label{eqn: sRP_components}
\eeqna
(The other components of $\tens{W}$ decouple to form a separate
equation set, with trivial dynamics.)  In the limit of fast chain
stretch relaxation $\taur \to 0$ we obtain the simpler
`non-stretching' RP model in which the trace $T=3$ and
\beqna
\dot{W}_{xy} &=& \dot{\gamma} \left[\wyy - \frac{2}{3} (1+\beta)\wxy^2\right]\;\;\;\;\;\;\; -\frac{1}{\taud}\wxy,+ D\partial_y^2 \wxy\nonumber\\
\dot{W}_{yy} &=& \frac{2}{3}\dot{\gamma}\left[\beta\wxy-(1+\beta)\wxy\wyy \right] - \frac{1}{\taud}(\wyy-1)+ D\partial_y^2\wyy. \quad \quad
\label{eqn: nRP_components}
\eeqna
\end{widetext}
For convenient shorthand we shall refer to this simpler non-stretching
form as the nRP model. We refer to the full `stretching' model of
Eqns.~\ref{eqn: sRP_components} as the sRP model.

For boundary conditions at the walls of the flow cell we assume no
slip and no permeation for the fluid velocity, and zero-gradient
$\partial_y W_{\alpha\beta}=0$ for every component $\alpha\beta$ of
the polymeric conformation tensor.

In what follows we consider the behaviour of the Rolie-poly model in the following two flow protocols:

\begin{itemize}

\item LAOStrain, with an imposed strain 
\beqn
\gamma(t)=\gamma_0\sin(\omega t),
\eeqn
to which corresponds the strain rate
\beqn
\gdot(t)=\gamma_0\omega\cos(\omega t)=\gdot_0\cos(\omega t).
\eeqn

\item LAOStress, with an imposed stress
\beqn
\Sigma(t)=\Sigma_0\sin(\omega t).
\eeqn

\end{itemize}

The model, flow geometry and protocol just described are characterised
by the following parameters: the polymer modulus $G$, the reptation
timescale $\taud$, the stretch relaxation timescale $\taur$, the CCR
parameters $\beta$ and $\delta$, the stress diffusivity $D$, the
solvent viscosity $\eta$, the gap size $L$, the frequency $\omega$ and
the amplitude $\gamma_0$ (for LAOStrain) or $\Sigma_0$ (for
LAOStress). We are free to choose units of mass, length and time,
thereby reducing the list by three: we work in units of length in
which the gap size $L = 1$, of time in which the reptation time $\taud
= 1$ and of mass (or actually stress) in which the polymer modulus
$G=1$.  We then set the value of the diffusion constant $D$ such that
the interface between the bands has a typical width $\ell =
\sqrt{D\taud}=2\times 10^{-2}L$, much smaller than the gap size. This
is the physically relevant regime for the macroscopic flow cells of
interest here, and we expect the results we report to be robust to
reducing $l$ further.  Following Ref.~\cite{Grahametal2003a} we set
$\delta = -\frac{1}{2}$.

Adimensional quantities remaining to be explored are then the model
parameters $\eta$, $\beta$ and (for the sRP model only) $\taur$; and
the protocol parameters $\omega$ and $\gamma_0$ or $\Sigma_0$. For
each set of model parameters we explore the whole plane of feasibly
accessible values of protocol parameters $\omega$ and $\gamma_0$ or
$\Sigma_0$.

Among the model parameters the CCR parameter has the range $0 \leq
\beta \leq 1$. Within this there is no current consensus as to its
precise value, and we shall therefore explore widely the full range
$0\to 1$. For the fluids of interest here the Newtonian viscosity is
typically much smaller than the zero shear viscosity of the
viscoelastic component, giving $\eta\ll 1$ in our units. Based on a
survey of the experimental data, a range of $10^{-7}$ to $10^{-3}$ was
suggested by Graham et al. in Ref.~\cite{Graham2013}.
Consistentwith comments made in Ref.~\cite{Agimelen2013} we find values
less than $10^{-5}$ unfeasible to explore numerically, due to a
resulting large separation of timescales between $\taud$ and $\eta/G$.
Therefore we adopt typical values $\eta=10^{-4}$ and $10^{-5}$.  Given
that that the susceptibility to shear banding increases with
decreasing $\eta$, we note that the levels of banding reported in what
follows are likely, if anything, to be an underestimate of what might
be observed experimentally. We return in our concluding remarks to
discuss this issue further.

We explore a wide range of values of the stretch relaxation time
$\taur$, or equivalently of the degree of entanglement
$Z=\taud/3\taur$: we consider $Z=1$ to $350$ for the sRP model (and
note that the nRP model has $Z\to\infty$ by definition).
Experimentally, values of $Z$ in the range of $50$ appear commonplace
and $100$ towards the upper end of what might currently be used
experimentally in nonlinear rheological studies.  One of the
objectives of this work is to provide a roadmap of values of $Z$ and
$\beta$ in which shear banding is expected to be observed, for typical
small values of $\eta$, in a sequence of LAOS protocols that scan
amplitude and frequency space.

\section{Calculation methods} 
\label{sec:methods}

In this section we outline the theoretical methods to be used
throughout the paper. In order to develop a generalised framework
encompassing both the nRP and sRP models, we combine all the relevant
dynamical variables (for any given model) into a state vector
$\vecv{s}$, with $\vecv{s}=(\wxy,\wyy)^T$ for the nRP model and
$\vecv{s}=(\wxy,\wyy,T)^T$ for the sRP model.  Alongside this we
define a projection vector $\vecv{p}$ of corresponding dimensionality
$d$, with $\vecv{p}=(1,0)$ for the nRP model and $\vecv{p}=(1,0,0)$
for sRP.

The total shear stress $\Sigma_{xy}=\Sigma$, from which we drop the
$xy$ subscript for notational brevity, is then given by
\beqn
\label{eqn: governing_eqn_force}
\Sigma(t) = G\vecv{p} \cdot \vecv{s}(y,t) + \eta \gdot(y,t),
\eeqn
and the viscoelastic constitutive equation has the generalised form
\beqn
\partial_{t\,}\vecv{s}(y,t) = \vecv{Q}(\vecv{s},\gdot) + D\partial_y^2\vecv{s}.
\label{eqn: governing_eqn_diffusive}
\eeqn
The dimensionality and functional form of $\vecv{Q}$ then specify the
particular constitutive model. In this way our generalised notation in
fact encompasses not only the nRP model (for which $d=2$) and sRP
model (for which $d=3$) but many more besides, including the Johnson
Segalman, Giesekus and Oldroyd B models~\cite{Larson1988}.

\subsection{Homogeneous base state}
\label{sec:base}

For any given applied flow our approach will be first to calculate the
fluid's response within the simplifying assumption that the
deformation must remain homogeneous across the cell. While this is an
artificial (and indeed incorrect) constraint in any regime where shear
banding is expected, it nonetheless forms an important starting point
for understanding the mechanism by which shear banding sets in.  (We
also note that most papers in the literature make this assumption
throughout, thereby disallowing any possibility of shear banding
altogether.)

Within this assumption of homogeneous flow, the response of the system
follows as the solution to the set of ordinary differential equations
\beqn
\label{eqn: governing_eqn_force_local}
\base{\Sigma}(t) = G\vecv{p} \cdot \base{\vecv{s}}(t) + \eta \base{\gdot}(t),
\eeqn
and
\beqn 
\dot{\base{\vecv{s}}}(t) = \vecv{Q}(\base{\vecv{s}},\base{\gdot}).
\label{eqn: governing_eqn_diffusive_local}
\eeqn
In these either $\base{\gdot}(t)$ or $\base{\Sigma}(t)$ is imposed, in
LAOStrain and LAOStress respectively, and the other dynamical
quantities are calculated numerically using an explicit Euler
algorithm~\cite{NumRecipes}. We use the `hat' notation to denote
that the state being considered is homogeneous.

\subsection{Linear stability analysis}
\label{sec:lsa}

Having calculated the behaviour of the fluid within the assumption
that the flow remains homogeneous, we now proceed to consider whether
this homogeneous `base state' flow will, at any point during an applied
oscillatory protocol, be unstable to the formation of shear bands. To
do so we add to the base state, for which we continue to use the hat
notation, heterogeneous perturbations of (initially) small amplitude:
\beqna
\Sigma(t) &=& \base{\Sigma}(t),\nonumber\\
\gdot(y,t)&=& \base{\gdot}(t) + \sum_{n=1}^\infty \delta \gdot_n(t) \cos(n\pi y/L),\nonumber\\
\vecv{s}(y,t) &=& \vecv{\base{s}}(t) + \sum_{n=1}^\infty \delta\vecv{s}_n(t) \cos(n\pi y/L).
\label{eqn: LSA}
\eeqna
Note that the total stress $\Sigma$ is not subject to heterogeneous
perturbations because the constraint of force balance decrees that it
must remain uniform across the gap, at least in a planar shear cell.
Substituting Eqns.~\ref{eqn: LSA} into Eqns.~\ref{eqn:
  governing_eqn_force} and~\ref{eqn: governing_eqn_diffusive}, and
expanding in successive powers of the magnitude of the small
perturbations $\delta{\gdot_n},\vecv{\delta s_n}$, we recover at
zeroth order Eqns.~\ref{eqn: governing_eqn_force_local} and~\ref{eqn:
  governing_eqn_diffusive_local} for the dynamics of the base state. At first order the heterogeneous perturbations obey
\beqna
\label{eqn: perturbation}
0&=&G\tens{p}\cdot \delta\vecv{s}_n(t)+\eta\delta\gdot_n(t),\nonumber\\
\dot{\delta\vecv{s}}_n &=& \tens{M}(t) \cdot \delta\vecv{s}_n + \tens{q}\delta{\gdot}_n,
\eeqna
in which  $\tens{M} =
\partial_{\vecv{s}\,}\vecv{Q}|_{\vecv{\base{s}},\base{\gdot}}-\tens{\delta}D(n\pi/L)^2$
and $\vecv{q}
= \partial_{\gdot}\vecv{Q}|_{\vecv{\base{s}},\base{\gdot}}$. Combining
these gives
\beqn
\label{eqn: one}
\dot{\delta\vecv{s}}_n = \tens{P}(t) \cdot \delta\vecv{s}_n,
\eeqn
with
\beqn
\tens{P}(t) = \tens{M}(t) - \frac{G}{\eta}\vecv{q}(t)\, \vecv{p}.
\label{eqn: two}
\eeqn
In any regime where the heterogeneity remains small, terms of second
order and above can be neglected.

To determine whether at any time $t$ during an imposed oscillatory
flow the heterogeneous perturbations
$\delta{\gdot}_n,\delta\vecv{s}_n(t)$ have positive rate of growth,
indicating linear instability of the underlying homogeneous base state
to the onset of shear banding, we consider first of all the
instantaneous sign of the eigenvalue $\eigen(t)$ of $\tens{P}(t)$ that
has the largest real part. A positive value of $\eigen(t)$ is clearly
suggestive that heterogeneous perturbations will be instantaneously
growing at that time $t$. We note, however, that the concept of a
time-dependent eigenvalue must be treated with caution.  In view of
this we cross check predictions made on the basis of the eigenvalue by
also directly numerically integrating the linearised Eqns.~\ref{eqn:
  one} using an explicit Euler algorithm. This allows us to determine
unambiguously whether the heterogeneous perturbations will be at any
instant growing (taking the system towards a banded state) or decaying
(restoring a homogeneous state), at the level of this linear
calculation.

In these linear stability calculations we neglect the diffusive term
in the viscoelastic constitutive equation, setting $D=0$. Reinstating
it would simply transform any eigenvalue $\eigen \to \eigen_n=\eigen -
D n^2\pi^2/L^2$ and provide a mechanism whereby any heterogeneity with
a wavelength of order the microscopic lengthscale $l$, or below,
diffusively decays. Accordingly the results of this linear calculation
only properly capture the dynamics of any heterogeneous perturbations
that have macroscopically large wavelengths, which are the ones of
interest in determining the initial formation of shear bands starting
from a homogeneous base state.

As a measure of the degree of flow heterogeneity at any time $t$ in
this linear calculation, we shall report in our results sections below
$\delta\gdot (t)$ normalised by the amplitude of the imposed
oscillation $\gdot_0$ in LAOStrain, or by $1+|\gdot(t)|$ in LAOStress,
where $\gdot(t)$ is the instantaneous value of the shear rate. (We
find numerically that bands tend to form in LAOStress when
$|\gdot(t)|\gg 1$. The additional 1 in the normalisation is used
simply to prevent the divergence of this measure when $\gdot(t)$ passes
through 0 in each half cycle.) Note that we no longer need to specify
the mode number $n$ for $\delta\gdot$, because within the assumption
$D=0$ just described, we are confining our attention to the limit of
long wavelength modes only and noting them all to have the same
dynamics, to within small corrections set by $D$.

\subsection{Full nonlinear simulation}
\label{sec:nonlinear}

While the linear analysis just described provides a calculationally
convenient method for determining whether shear banding will arise in
any given oscillatory measurement, enabling us to quickly build up an
overall roadmap of parameter space, it cannot predict the detailed
dynamics of the shear bands once the amplitude of heterogeneity has
grown sufficiently large that nonlinear effects are no longer
negligible. Therefore in what follows we shall also perform full
nonlinear simulations of the model's spatio-temporal dynamics by
directly integrating the full model Eqns.~\ref{eqn:
  governing_eqn_force} and~\ref{eqn: governing_eqn_diffusive} using a
Crank-Nicolson algorithm \cite{NumRecipes}, with the system's
state discretised on a grid of $J$ values of the spatial coordinate
$y$, checked in all cases for convergence with respect to increasing
the number of grid points.

As a measure of the degree of shear banding at any time $t$ in this
nonlinear calculation we report the difference between the maximum and
minimum values of the shear rate across the cell:
\beqn
\Delta_{\gdot}(t) = \frac{1}{N}\Big[|\gdot_{\rm max}(t) - \gdot_{\rm min}(t)|\Big],
\label{eqn: dob}
\eeqn
again normalised depending upon the employed protocol, by $N$, where $N$ is the amplitude of the imposed oscillation $\gdot_0$ in LAOStrain, and $1+|\gdot(t)|$ in LAOStress.

\subsection{Seeding the heterogeneity}
\label{sec:seed}

When integrating the model equations to determine the time evolution
of any flow heterogeneity, whether linearised or in their full
nonlinear form, we must also specify the way in which whatever
heterogeneous perturbations that are the precursor to the formation of
shear bands are seeded initially. Candidates include any residual
heterogeneity left in the fluid by the initial procedure of sample
preparation; imperfections in the alignment of the rheometer plates;
true thermal noise with an amplitude set by $k_{\rm B}T$; and
rheometer curvature in cone-and-plate or cylindrical Couette devices.
We consider in particular the last of these because it is likely to be
the dominant source of heterogeneity in commonly used flow cells,
which typically have a curvature of about $10\%$.

While modelling the full effects of curvature is a complicated task,
its dominant consequence can be captured simply by including a slight
heterogeneity in the total stress field. (The assumption made above of
a uniform stress across the gap only holds in an idealised planar
device.) Accordingly we set $\Sigma(t)\to\Sigma(t)\left[1+q
  h(y)\right]$ where $q$ sets the amplitude of the curvature and
$h(y)$ is a function with an amplitude of $O(1)$ that prescribes its
spatial dependence. The detailed form of $h(y)$ will differ from
device to device: for example in a cylindrical Couette it is known to
have a $1/r^2$ dependence, where $r$ is the radial coordinate.
However, the aim here is not to model any particular device geometry
in detail, but simply to capture the dominant effect of curvature in
seeding the flow heterogeneity. Accordingly we set
$h(y)=\cos(\pi/L)$ which is the lowest Fourier mode to fit into the
simulation cell while still obeying the boundary conditions at the
walls.

\section{Shear banding in other time dependent protocols}
\label{sec:recap}

As a preamble to presenting our results for shear banding in
oscillatory flow protocols in the next two sections below, we first
briefly collect together criteria derived in previous work for linear
instability to the formation of shear bands in simpler time-dependent
protocols: slow shear rate sweep, fast shear startup, and step stress.

\subsection{Slow shear rate sweep}
\label{sec:recapSweep}

A common experimental protocol consists of slowly sweeping the shear
rate $\gdot$ upwards (or downwards) in order to measure a fluid's
(quasi) steady state flow curve. In this protocol the criterion for
linear instability to the onset of shear banding, given a base state
of initially homogeneous shear flow, has long been known to
be~\cite{Yerushalmi1970}
\beqn
\label{eqn:criterionSteady}
\frac{\partial\Sigma}{\partial\gdot} < 0.
\eeqn

\subsection{Fast shear startup}
\label{sec:recapStartup}

Another common experimental protocol consists of taking a sample of
fluid that is initially at rest and with any residual stresses well
relaxed, then suddenly jumping the strain rate from zero to some
constant value such that $\gdot(t)=\gdot_0\Theta(t)$, where
$\Theta(t)$ is the Heaviside function. Commonly measured in response
to this applied flow is the time-dependent stress signal $\Sigma(t)$
as it evolves towards its eventual steady state value, for that
particular applied shear rate, on the fluid's flow curve. This
evolution typically has the form of an initial elastic regime with
$\Sigma\approx G\gamma$ while the strain $\gamma$ remains small,
followed by an overshoot in the stress at a strain of $O(1)$, then a
decline to the final steady state stress on the flow curve. In
Ref.~\cite{Moorcroft2013,Moorcroft2014,Adamsetal2011}
we gave evidence that the presence of an overshoot in this stress
startup signal is generically indicative of a strong tendency to form
shear bands, at least transiently. These bands may, or may not, then
persist for as long as the shear remains applied, according to whether
or not the underlying constitutive curve of stress as a function of
strain rate is non-monotonic.

Such behaviour is to be expected intuitively. Consider a shear startup
run performed at a high enough strain rate that the material's
response is initially elastic, with the stress startup signal
depending only on the accumulated strain $\gamma=\gdot t$ and not
separately on the strain rate $\gdot$. The decline in stress following
an overshoot in the stress startup signal corresponds to a negative
derivative
\beqn
\label{eqn:criterionSimpleElastic}
\frac{\partial \Sigma}{\partial\gamma}<0.
\eeqn
This clearly has the same form as~(\ref{eqn:criterionSteady}) above,
with the strain rate now replaced by the strain.  As such it is the
criterion that we might intuitively expect for the onset of strain
bands in a nonlinear elastic solid, following the early intuition of
Ref.~\cite{Marrucci1983}

In close analogy to this intuitive expectation, for a complex fluid
subject to a fast, elastically dominated startup the criterion for the
onset of banding was shown in Ref.~\cite{Moorcroft2013} to be
that the stress signal $\Sigma(\gamma=\gdot t)$ of the initially
homogeneous startup flow obeys
\beqn 
\label{eqn:criterionStartup}
-\textrm{tr}\tens{M} \frac{\partial \Sigma}{\partial\gamma} +
\gdot\frac{\partial^2\Sigma}{\partial\gamma^2} < 0,
\eeqn
where $\textrm{tr}\tens{M}<0$ in this startup protocol.  This result
holds exactly for any model whose equations are of the generalised
form in Sec.~\ref{sec:methods} above, and have only two relevant
dynamical variables, $d=2$. (Recall that for the nRP model these two
variables are the shear stress $W_{xy}$ and one component of normal
stress $W_{yy}$, in units in which the polymer modulus $G=1$.)  The
criterion~(\ref{eqn:criterionStartup}) closely resembles the simpler
form~(\ref{eqn:criterionSimpleElastic}) motivated intuitively above,
with an additional term informed by the curvature in the signal of
stress as a function of strain. The effect of this additional term is
to trigger the onset of banding just {\em before} overshoot, as the
stress startup signal starts to curve downwards from its initial
regime of linear elastic response.

What this criterion tells us is that the presence of an overshoot in the
stress signal of an underlying base state of initially homogeneous shear startup acts as a causative trigger for the formation of shear bands. A common misconception is that instead it is the onset of shear
banding that causes the stress drop.  While it is true that the onset
of banding may reduce the stress further compared to that expected on
the basis of a homogeneous calculation, we emphasise that the
direction of mechanistic causality here is that the stress drop
following overshoot causes shear banding and not (primarily) vice
versa.

With criterion (\ref{eqn:criterionStartup}) in mind, theorists should
be alert that any model predicting startup stress overshoot in a
calculation in which the flow is artificially constrained to remain
homogeneous is likely to further predict the formation of shear bands
in a full heterogeneous calculation that allows bands to form. Likewise
experimentalists should be alert that any observations of stress
overshoot in shear startup is strongly suggestive of the presence of
banding in the material's flow profile.

In Ref.~\cite{Moorcroft2014} the analytically derived
criterion~(\ref{eqn:criterionStartup}) was confirmed numerically for
fast shear startup in the nRP model, where it should indeed apply
exactly due to the presence of just $d=2$ relevant dynamical variables
$W_{xy}$ and $W_{yy}$ in that model. It was also shown to apply to
good approximation in the sRP model, for which $d=3$, for strain rates
lower than the inverse stretch relaxation time (where the dynamics of
the sRP model indeed well approximate those of the nRP model).

Banding associated with startup stress overshoot has also been
demonstrated in several numerical studies of soft glassy materials
(SGMs)~\cite{Moorcroft2011,Fielding2014,Manningetal2009a}.
(The term SGM is used to describe a broad class of materials including
foams, emulsions, colloids, surfactant onion phases and microgels, all
of which show structural disorder, metastability, a yield stress, and
often also rheological ageing below the yield stress.)  In these soft
glasses, however, it should be noted that the decrease in stress
following the startup overshoot arises from increasing plasticity
rather than falling elasticity. This makes it more difficult to derive
an analytical criterion analogous to~(\ref{eqn:criterionStartup}).
Accordingly the theoretical evidence for shear banding following
startup overshoot in these soft glasses, while very convincing,
remains primarily numerical to date.

Consistent with these theoretical predictions, experimental
observations of banding associated with startup stress overshoot are
widespread: in wormlike micellar surfactants~\cite{Hu2008,
  Wangetal2008c}, polymers~\cite{Wangetal2008a, Wangetal2008b,
  Wangetal2009a, Wangetal2009b, Boukany2009a, Wangetal2003a,
  Huetal2007a, Wangetal2009d}, carbopol
gels~\cite{Divoux2011a,Divoux2010} and Laponite clay
suspensions~\cite{Martin2012}. Nonetheless, we also note other
  studies of polymer solutions~\cite{Li2015} where stress overshoot is
  seen without observable banding. It would be particularly
  interesting to see further experimental work on polymeric fluids to
  delineate more fully the regimes, for example of entanglement number
  and degree of polydispersity, in which banding arises with
  sufficient amplitude to be observed experimentally.

\subsection{Step stress}
\label{sec:recapCreep}

Besides the strain-controlled protocols just described, a fluid's
rheological behaviour can also be probed under conditions of imposed
stress.  In a step stress experiment, an initially well relaxed fluid
is suddenly subject to the switch-on of a shear stress $\Sigma_0$ that
is held constant thereafter, such that $\Sigma(t)=\Theta(t)\Sigma_0$.
Commonly measured in response to this applied stress is the material's
creep curve, $\gamma(t)$, or the temporal derivative of this,
$\gdot(t)$. In Ref.~\cite{Moorcroft2013} the criterion for
linear instability to the formation of shear bands, starting from a
state of initially homogeneous creep shear response, was shown to be
that
\beqn
\frac{\partial^2\gdot}{\partial t^2}/\frac{\partial\gdot}{\partial t}>0.
\eeqn
This tells us that shear banding should be expected in any step stress
experiment in which the differentiated creep response curve
simultaneously curves upwards and slopes upwards. (Indeed it should
also be expected in any experiment where that response function
simultaneously curves downwards and slopes downwards, though we do not
know of any instances of such behaviour.) This prediction has been
confirmed numerically in the Rolie-poly model of polymers and wormlike
micelles~\cite{Moorcroft2014}, as well as in the soft glassy
rheology model of foams, dense emulsions, microgels, {\it
  etc}~\cite{Fielding2014}.

\begin{figure}[tbp]
\includegraphics[width=9.0cm]{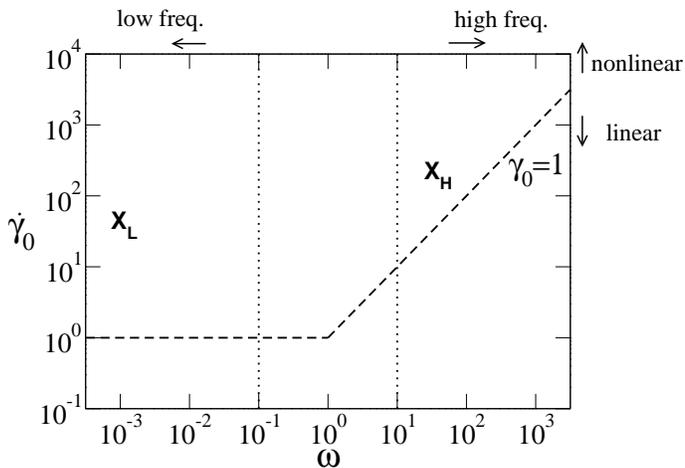}
\caption{LAOStrain: sketch of regions of shear rate amplitude and
  frequency space in which we expect limiting low frequency `viscous'
  and high frequency `elastic' behaviours, and regimes of linear and
  nonlinear response. LAOStrain runs at the locations marked $X_L$ and
  $X_H$ are explored in Figs.~\ref{fig:nonmon} and~\ref{fig:mon} for
  the nRP model with non-monotonic and monotonic underlying
  constitutive curve respectively.}
\label{fig:sketch}
\end{figure}

Experimentally, shear banding associated with a simultaneously
upwardly curving and upwardly sloping differentiated creep response
curve has indeed been seen in in entangled
polymers~\cite{Wangetal2009a,Huetal2007a,Wangetal2003a, Hu2008,
  Wangetal2008c, Hu2005, Hu2010, Wangetal2009d}, carbopol
microgels~\cite{Divoux2011} and carbon black
suspensions~\cite{Gibaud2010}.

\section{Large amplitude oscillatory strain}
\label{sec:LAOStrain}

We now consider shear banding in the time-dependent strain-imposed
oscillatory protocol of LAOStrain. Here a sample of fluid, initially
well relaxed at time $t=0$, is subject for times $t>0$ to a strain of
the form
\beqn
\gamma(t)=\gamma_0\sin(\omega t),
\eeqn
to which corresponds the strain rate
\beqn
\gdot(t)=\gamma_0\omega\cos(\omega t)=\gdot_0\cos(\omega t).
\eeqn
After an initial transient, once many cycles have been executed, the
response of the system is expected to attain a state that is
time-translationally invariant from cycle to cycle, $t\to
t+2\pi/\omega$. All the results presented below are in this long-time
regime, usually for the $N=20$th cycle after the flow
commenced. The dependence of the stress on the cycle number was carefully studied in wormlike micelles in Ref.~\cite{Fujii2015}.

To characterise any given applied LAOStrain we must clearly specify
two quantities: the strain amplitude and the frequency
$(\gamma_0,\omega)$, or alternatively the strain rate amplitude and
the frequency $(\gdot_0,\omega)$, where $\gdot_0=\gamma_0\omega$. In
what follows we usually choose the latter pairing $(\gdot_0,\omega)$.
Any given LAOStrain experiment is then represented by its location in
that plane of $\gdot_0$ and $\omega$. See Fig.~\ref{fig:sketch}.

In any experiment where the applied strain rate remains small,
$\gdot_0 \ll 1$, a regime of linear response is expected.  (Recall
that in dimensional form this condition corresponds to $\gdot_0\taud
\ll 1$.)  But even in an experiment where the strain rate does not
remain small, linear response can nonetheless still be expected if the
overall applied strain remains small, $\gamma_0 \ll 1$. Accordingly,
linear response should obtain in the region below the long-dashed line
marked in Fig.~\ref{fig:sketch}. Because shear banding is an
inherently nonlinear phenomenon, we expect the interesting region of
this $(\gdot_0,\omega)$ plane from our viewpoint to be in the
nonlinear regime, above the long-dashed line, and we focus our
attention mostly on this in what follows.

Besides considering whether any given applied LAOStrain will result in
linear or nonlinear response, also relevant is the characteristic
timescale $1/\omega$ of the oscillation compared to the fluid's
intrinsic terminal relaxation timescale $\taud=1$. For low
frequencies $\omega\ll 1$, to the left of the leftmost dotted line in
Fig.~\ref{fig:sketch}, we expect the material's reconfiguration
dynamics to keep pace with the applied deformation. This will lead to
quasi steady state response in which the stress slowly sweeps up
and down the steady state flow curve as the shear rate varies through
a cycle. In contrast for high frequencies $\omega\gg 1$, to the right
of the rightmost dotted line in Fig.~\ref{fig:sketch}, the material's
relaxation dynamics cannot keep pace with the applied deformation and
we expect elastic-like response.

\begin{figure}[tbp]
\includegraphics[width=8.0cm]{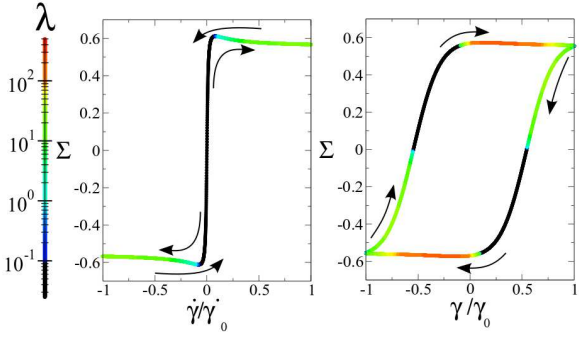}
\caption{LAOStrain in the nRP model with a non-monotonic underlying
  constitutive curve. Model parameters $\beta=0.4$, $\eta=10^{-5}$.
  {\bf Left:} Viscous Lissajous-Bowditch
  figure shows stress $\Sigma$ versus strain rate $\gdot$ for an
  imposed frequency and strain rate $(\omega,\gdot_0)=(0.001,50.0)$
  marked as $X_L$ in the low frequency regime of
  Fig.~\ref{fig:sketch}.  {\bf Right:} Elastic Lissajous-Bowditch figure shows
  stress $\Sigma$ versus strain $\gamma$ for an imposed frequency and
  strain rate $(\omega,\gdot_0)=(31.6,200.0)$ marked as $X_H$ in the
  high frequency regime of Fig.~\ref{fig:sketch}.  Colourscale shows
  eigenvalue.}
\label{fig:nonmon}
\end{figure}

\begin{figure}[tbp]
\includegraphics[width=8.0cm]{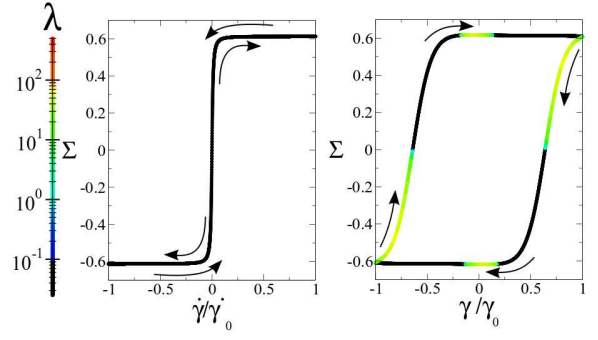}
\caption{As in Fig.~\ref{fig:nonmon}, but for a value of the CCR
  parameter $\beta=1.0$, for which the underlying homogeneous
  constitutive curve is monotonic.}
\label{fig:mon}
\end{figure}

We illustrate these two limiting regimes by studying the response of
the nRP model to an imposed LAOStrain at each of the two locations
marked $X_L$ and $X_H$ in Fig.~\ref{fig:sketch}.  For simplicity, for
the moment, we artificially constrain the flow to remain homogeneous
and confine ourselves to calculating the uniform `base state' as
outlined in Sec.~\ref{sec:base}. The results are shown in
Fig.~\ref{fig:nonmon} for the nRP model with parameters for which the
underlying constitutive curve is non-monotonic, such that (in any
heterogeneous calculation) the fluid would show shear banding under
conditions of steady applied shear.  Fig.~\ref{fig:mon} shows results
with model parameters for which the constitutive curve is monotonic,
such that no banding would be expected in steady shear flow.

The left panels of Figs.~\ref{fig:nonmon} and~\ref{fig:mon} contain
results for the low frequency oscillation marked $X_L$ in
Fig.~\ref{fig:sketch}. Here we choose to plot the stress response
$\Sigma(t)$ in a Lissajous-Bowditch figure as a parametric function of
the time-varying imposed strain rate $\gdot(t)$, consistent with the
expectation of fluid-like response in this low-frequency regime.
(Throughout the paper we shall describe such a plot of stress versus strain rate as being in the
`viscous' representation.) As can be seen, in each case the fluid
indeed tracks up and down its (quasi) steady state homogeneous
constitutive curve $\Sigma(\gdot)$ in the range $-\gdot_0 < \gdot <
\gdot_0$.  For any set of model parameters, several of these LAOStrain
response curves $\Sigma(\gdot)$ collected together for different
$\gdot_0$ and low frequency $\omega$ would all collapse onto this
master constitutive curve.

Also shown by the colour scale in the left panels of
Figs.~\ref{fig:nonmon} and~\ref{fig:mon} is the eigenvalue as
introduced in Sec.~\ref{sec:lsa}. Recall that a positive eigenvalue at
any point in the cycle strongly suggests that the homogeneous base
state is linearly unstable to the development of shear banding at that
point in the cycle.  (In any region where this scale shows black the
eigenvalue is either negative, or so weakly positive as to cause only
negligible banding growth.) As expected, a regime of instability is
indeed seen in Fig.~\ref{fig:nonmon}, in the region
where the constitutive curve has negative slope,
\beqn
\frac{\partial\Sigma}{\partial\gdot} <0.
\eeqn
For a fluid with a monotonic constitutive curve, no instability is
observed at this low frequency (Fig.~\ref{fig:mon}, left).

The corresponding results for the high frequency run marked $X_H$ in
Fig.~\ref{fig:sketch} are shown in the right panels of
Figs.~\ref{fig:nonmon} and~\ref{fig:mon}. Here we choose to plot the
stress response $\Sigma(t)$ in a Lissajous-Bowditch figure as a
parametric function of the time-varying strain $\gamma(t)$, in the
so-called `elastic' representation. Indeed, just as in the low
frequency regime the material behaved as a viscous fluid with the
stress response falling onto the steady state master constitutive
curve in the viscous representation $\Sigma(\gdot)$, for a high
frequency cycle we might instead expect a regime of elastic response
in which only the accumulated strain is important, and not
(separately) the strain rate, giving a master response curve of stress
versus strain, $\Sigma(\gamma)$.

We might further have intuitively expected this curve to be the same
as that obtained in a fast shear startup from rest, with (in the
positive strain part of the cycle) elastic response $\Sigma\approx G\gamma$
at low strain $\gamma\ll 1$, followed by stress overshoot at a typical
strain $\gamma=+O(1)$, then decline towards a constant stress at
larger strains (with the symmetric curve in the negative-strain part
of the cycle, such that $\Sigma\to-\Sigma$ for $\gamma\to-\gamma$). In
other words, in LAOStrain at high frequency we might have expected the
system to continuously explore its elastic shear startup curve
$\Sigma(\gamma)$ between $\gamma=-\gamma_0$ and $\gamma=+\gamma_0$.

However, this intuition is not met in a straightforward way. In the
right panels of Fig.~\ref{fig:nonmon} and~\ref{fig:mon} we observe
instead an open cycle that is explored in a clockwise sense as time
proceeds through an oscillation: the stress transits the upper part of
the loop (from bottom left to top right) in the forward part of the
cycle as the strain increases from left to right, and the
symmetry-related lower part of the loop in the backward part, where
the strain decreases from right to left.

This can be understood as follows. For any LAOStrain run at high
frequency $\omega \gg 1$ but in the linear regime with strain
amplitude $\gamma_0\ll 1$, we do indeed find the stress response to
fall onto a closed master curve $\Sigma(\gamma)$, which also
corresponds to that obtained in a fast stress startup from rest, with
linear elastic response $\Sigma\approx G\gamma$. (Data not shown.) In
contrast, for amplitudes $\gamma_0 > 1$ the system only explores this
startup-from-rest curve in the first half of the {\em first} cycle
after the inception of flow. (This has the usual form, with elastic
response for small strains, stress overshoot at a strain
$\gamma=O(1)$, then decline to a constant stress.)  In the second half
of the cycle, when the strain rate reverses and the strain decreases,
the stress response departs from the startup-from-rest curve. With
hindsight this is in fact obvious: as this backward shear part of the
cycle commences the initial condition is not that of a well-relaxed
fluid, but one that has just suffered a large forward strain.

The same is true for the next forward half cycle: its initial
condition is that of a fluid that has just suffered a large negative
strain, corresponding to the lower left point in the right panel of
Figs.~\ref{fig:nonmon} or~\ref{fig:mon}. Starting from that initial
condition the stress evolution nonetheless thereafter resembles that
of a fast startup, with an initial fast rise followed by an overshoot
then decline to constant stress, before doing the same in reverse
(with a symmetry-related `negative overshoot') during the next half
cycle, giving the open curves as described. Associated with this
overshoot in each half cycle is a positive eigenvalue indicating
instability to the onset of shear banding. Importantly, we note that
this arises even in the case of a monotonic underlying constitutive
curve (Fig.~\ref{fig:mon}, right), and therefore even in a fluid that would
not display steady state banding under a steadily applied shear flow.
It is the counterpart for LAOStrain of the `elastic' banding triggered
by stress overshoot in a fast shear startup from rest, as explored
previously in
Ref.~\cite{Moorcroft2013,Moorcroft2014,Adamsetal2011}.

\begin{figure}[tbp]
\includegraphics[width=9.0cm]{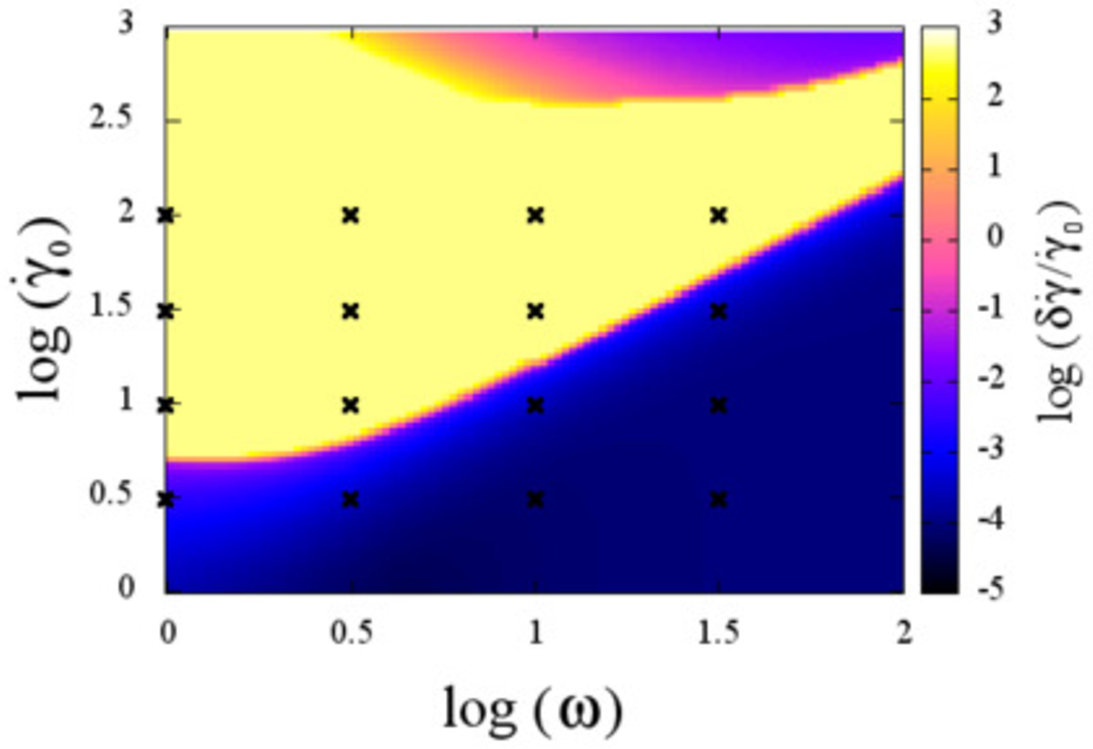}
\caption{Colour map of the normalised degree of shear banding for the
  nRP model with a non-monotonic constitutive curve. Each point in this
  $\gdot_0,\omega$ plane corresponds to a particular LAOStrain run
  with strain rate amplitude $\gdot_0$ and frequency $\omega$.  For
  computational efficiency, these calculations are performed by
  integrating the linearised equations in Sec.~\ref{sec:lsa}. Reported
  is the maximum degree of banding that occurs at any point in the
  cycle, after many cycles.  Model parameters: $\beta=0.4$,
  $\eta=10^{-5}$.  Cell curvature $q=10^{-4}$. Crosses indicate the
  grid of values of $\gdot_{0}$ and $\omega$ in Pipkin diagram of
  Fig.~\ref{fig:PipkinNonMon}.}
\label{fig:pinPointNonMon}
\end{figure}

\begin{figure}[tbp]
\includegraphics[width=9.0cm]{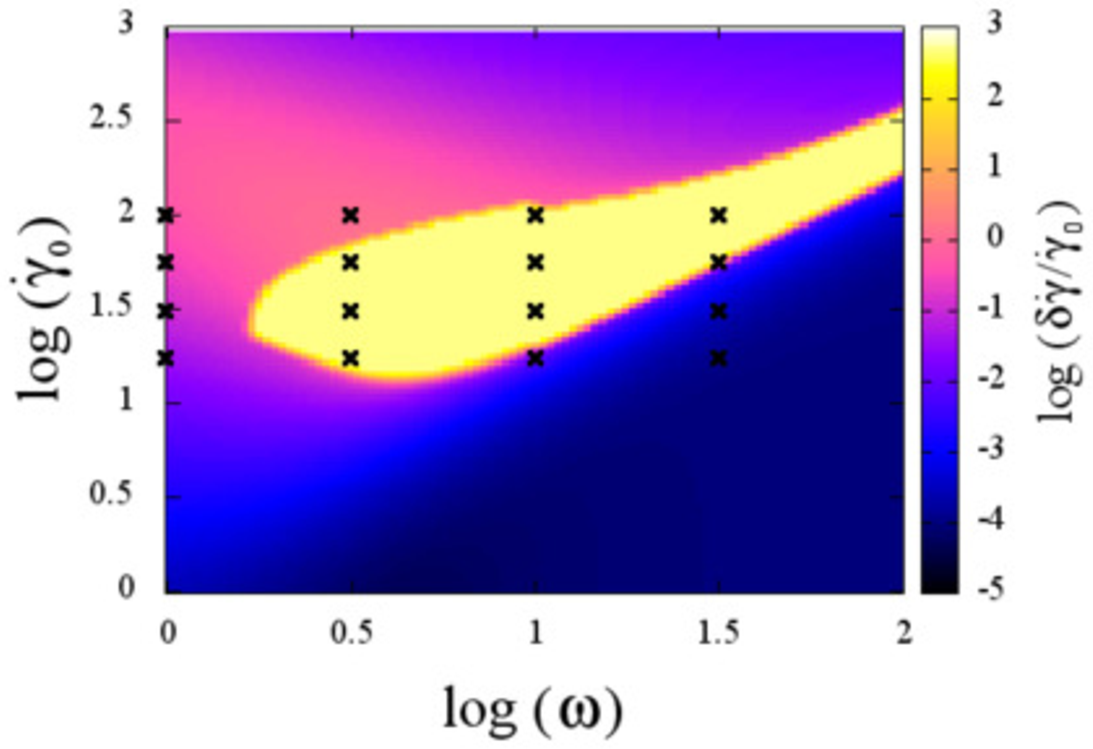}
\caption{As in Fig.~\ref{fig:pinPointNonMon}, but with a CCR parameter
  $\beta=1.0$, for which the fluid has a monotonic underlying
  constitutive curve.  Crosses indicate the grid of values of
  $\gdot_{0}$ and $\omega$ used in the Pipkin diagram of
  Fig.~\ref{fig:PipkinMon}.}
\label{fig:pinPointMon}
\end{figure}

\begin{figure}[tp!]
\includegraphics[width=9cm]{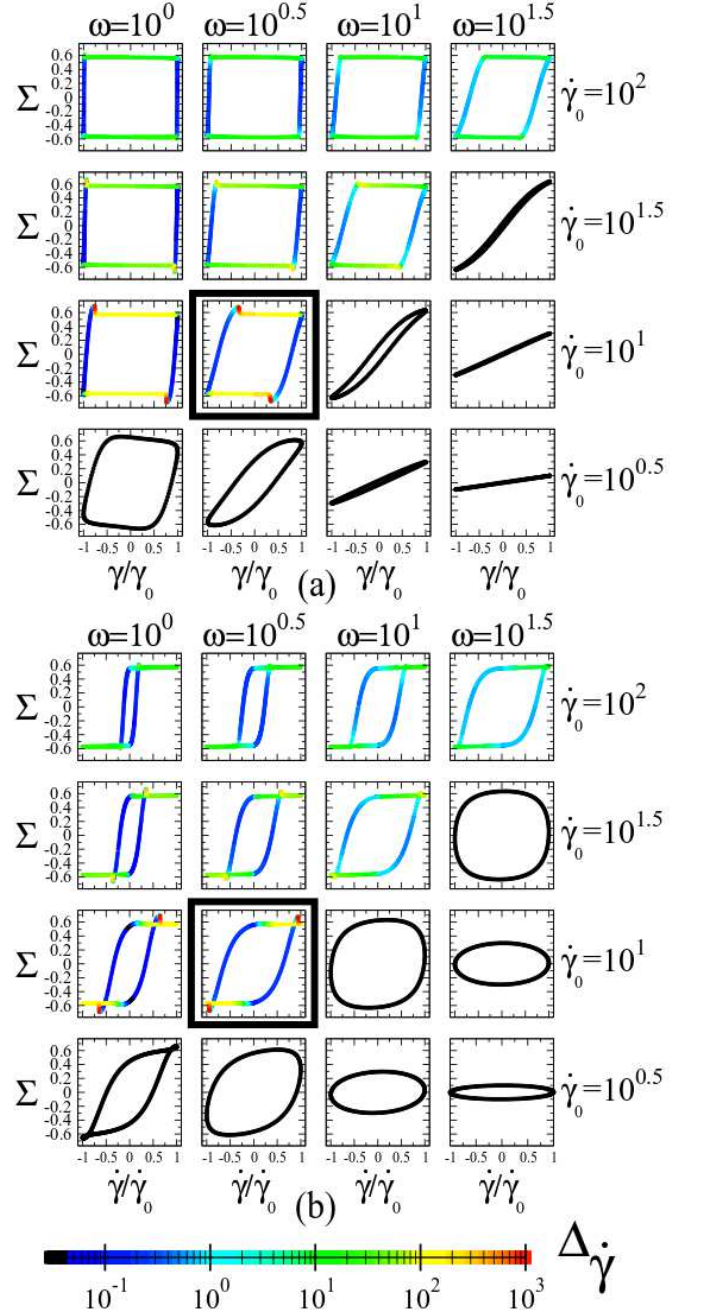}
\caption{Lissajous-Bowditch curves in LAOStrain for the nRP model with
  a non-monotonic constitutive curve. Results are shown in the elastic
  representation in (a), and the viscous representation in
  (b). Columns of fixed frequency $\omega$ and rows of fixed
  strain-rate amplitude $\gdot_0$ are labeled at the top and
  right-hand side. Colourscale shows the time-dependent degree of
  shear banding.  Model parameters: $\beta=0.4, \eta=10^{-5}$. Cell
  curvature: $q=10^{-4}$. Number of numerical grid points $J=512$. A detailed portrait of the run outlined by
  the thicker box is shown in Fig.~\ref{fig:portraitNonMon}.}
\label{fig:PipkinNonMon}
\end{figure}

\begin{figure}[tp!]
\includegraphics[width=9cm]{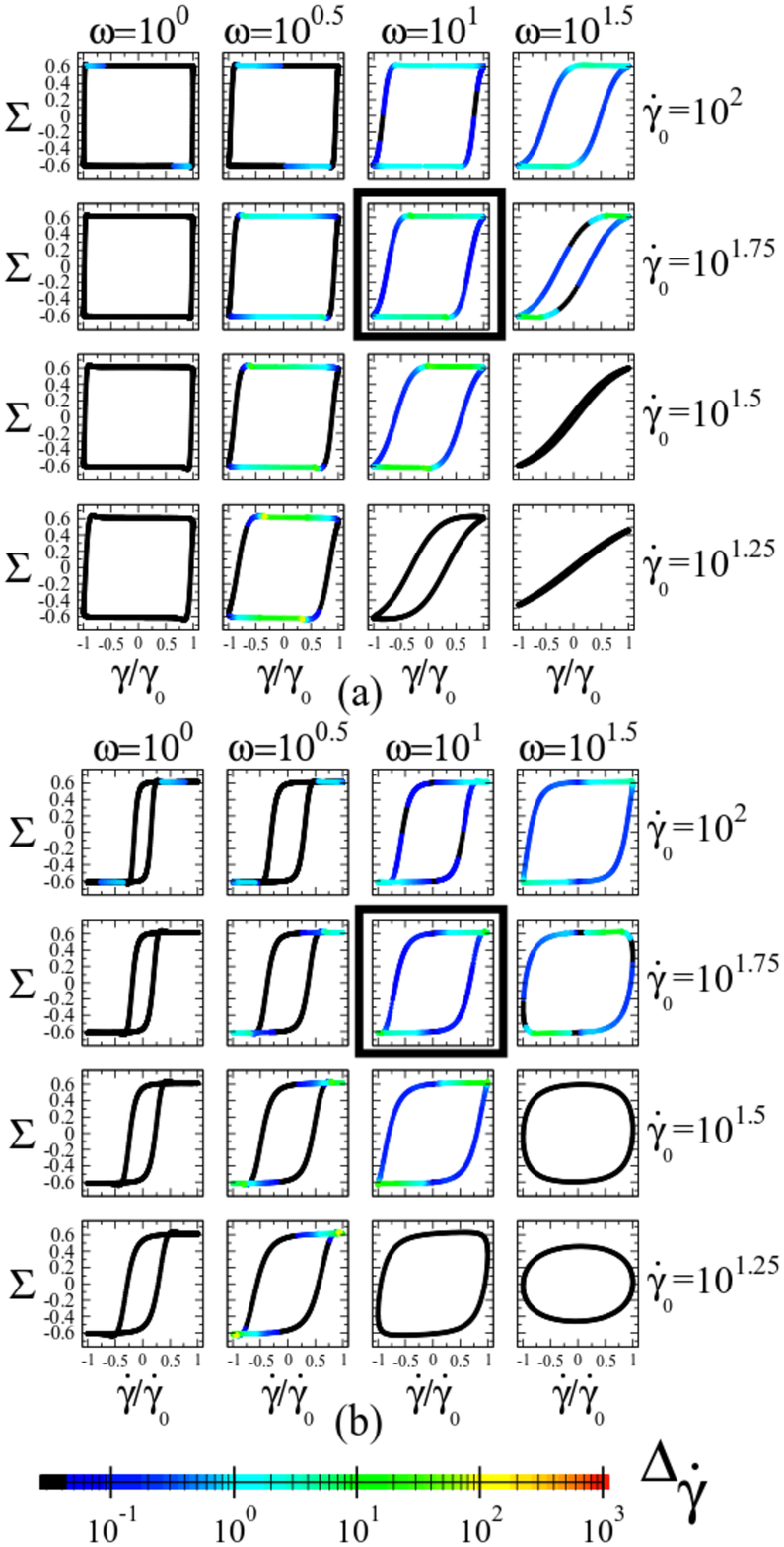}
\caption{As in Fig.~\ref{fig:PipkinNonMon} but for a value of the CCR
  parameter $\beta=1.0$, for which the fluid's underlying constitutive
  curve is monotonic. Number of numerical grid points $J=512$. A detailed portrait of the run outlined by the
  thicker box is shown in Fig.~\ref{fig:portraitMon}.}
\label{fig:PipkinMon}
\end{figure}

\begin{figure}[tbp]
\includegraphics[width=8cm]{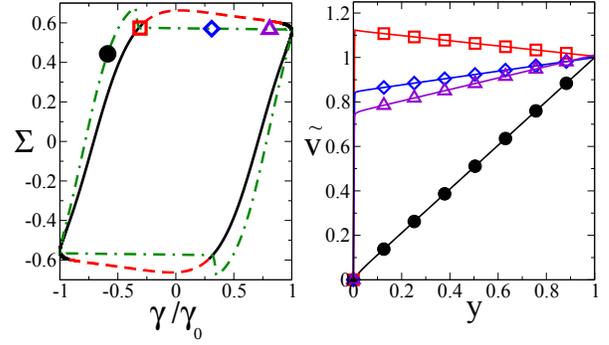}
\caption{LAOStrain in the nRP model with a non-monotonic constitutive
  curve. Strain rate amplitude $\gdot_{0}=10.0$ and frequency
  $\omega=3.16$. Model parameters $\beta=0.4, \eta=10^{-5}$.  Cell
  curvature $q=10^{-4}$. Number of numerical grid points $J=1024$.
  {\bf Left:} stress response in the elastic representation. Solid
  black and red-dashed line: calculation in which the flow is
  constrained to be homogeneous.  Red-dashed region indicates a
  positive eigenvalue showing instability to the onset of shear
  banding. Green dot-dashed line: stress response in a full nonlinear
  simulation that allows banding. {\bf Right:} Velocity profiles
  corresponding to stages in the cycle indicated by matching symbols
  in the left panel. Each profile is normalised by the speed of the
  moving plate.}
\label{fig:portraitNonMon}
\end{figure}

Indeed, following the calculation first set out in
Ref.~\cite{Moorcroft2013}, it is straightforward to show that
the condition for a linear instability to banding in this `elastic'
high frequency regime $\omega\gg 1$ is the same as in fast shear
startup:
\beqn 
\label{eqn:criterionElastic}
-\textrm{tr}\tens{M} \frac{\partial \Sigma}{\partial\gamma} +
\gdot\frac{\partial^2\Sigma}{\partial\gamma^2} < 0.
\eeqn
As already discussed, this gives a window of instability setting in
just before the stress overshoot (or negative overshoot) in each half
cycle in the right panels of Fig.~\ref{fig:nonmon} and~\ref{fig:mon}
due to (in the positive $\gdot$ part of the cycle in which the stress
transits from bottom left to top right) the negatively sloping and
curving $\Sigma(\gamma)$. An analogous statement applies in the other
part of the cycle, with the appropriate sign reversals. Note that
these overshoots are sufficiently weak as to be difficult to resolve
by eye on the scale of Figs.~\ref{fig:nonmon} and~\ref{fig:mon}.

Interestingly,~(\ref{eqn:criterionElastic}) also predicts a region of
(weaker) instability immediately after the reversal of strain in each
half cycle, as also seen in the right panels of Figs.~\ref{fig:nonmon}
and~\ref{fig:mon}.  Analytical considerations show that this
additional regime of instability is not driven by any negative slope or
curvature in $\Sigma(\gamma)$, but instead arises from a change in
sign of $\textrm{\tens{M}}$.  This instability has no counterpart that
we know of in shear startup. Its associated eigenvector is dominated
by the normal stress component $W_{yy}$ rather than the strain rate or
shear stress. Heterogeneity in this quantity could feasibly be
accessed in birefringence experiments.

The results of Figs.~\ref{fig:nonmon} and~\ref{fig:mon} can be
summarised as follows. At low frequencies the system sweeps slowly up
and down the underlying constitutive curve $\Sigma(\gdot)$ as the
shear rate varies through the cycle. If that curve is non-monotonic,
this homogeneous base state is unstable to shear banding in the region
of negative constitutive slope, $d\Sigma/d\gdot<0$. At high
frequencies the system instead executes a process reminiscent of
elastic shear startup in each half cycle, but with an initial
condition corresponding to the state left by the shear of opposite
sense in the previous half cycle. Associated with this is a stress
overshoot in each half cycle, giving instability to elastic shear
banding. Crucially, this elastic instability occurs whether or not the
underlying constitutive curve is non-monotonic or monotonic, and
therefore whether or not the fluid would shear band in steady shear.

From a practical experimental viewpoint it is important to note that,
whereas in a single shear startup run these `elastic' strain bands
would form transiently then heal back to homogeneous flow (unless the
sample has a non-monotonic underlying constitutive curve and so also
bands in steady state), in LAOStrain the banding will recur in each
half cycle and so be potentially easier to access experimentally.
Any time-averaging measurement should of course only take data in the
forward part of each cycle, to avoid averaging to zero over the
cycle.

Having explored in Figs.~\ref{fig:nonmon} and~\ref{fig:mon} the
tendency to form shear bands in two particular LAOStrain runs (one in
the limit of low frequency, $X_L$ in Fig.~\ref{fig:sketch}, and one in
the limit of high frequency, $X_H$), we now explore the full
$(\gdot_0,\omega)$ plane of Fig.~\ref{fig:sketch} by showing in
Figs.~\ref{fig:pinPointNonMon} and~\ref{fig:pinPointMon} colour maps of
the extent of shear banding across this plane.  Recall that each
pinpoint in this plane corresponds to a single LAOS experiment with
strain rate amplitude $\gdot_0$ and frequency $\omega$.  To build up
these colour maps, we sweep over a grid of 100x100
values of $\gdot_0,\omega$ and execute a LAOStrain run at each point.
Solving the model's full nonlinear dynamics on such a dense grid would
be unfeasibly time-consuming computationally. Therefore at each
$\gdot_0,\omega$ we instead integrate the linearised equations set out
in Sec.~\ref{sec:lsa}. In each such run we record the degree of
banding $\delta\gdot$, maximised over the cycle after many cycles.  It
is this quantity, normalised by the maximum strain rate amplitude $\gdot_{0}$, that is represented by the colourscale in
Figs.~\ref{fig:pinPointNonMon} and~\ref{fig:pinPointMon}.

Fig.~\ref{fig:pinPointNonMon} pertains to the nRP model with model
parameters for which the underlying constitutive curve is
non-monotonic. As expected, significant banding (bright/yellow region)
is observed even in the limit of low frequency $\omega\to 0$ for
strain rate amplitudes $\gdot_0$ exceeding the onset of negative slope
in the underlying constitutive curve. This region of banding is the
direct (and relatively trivial) analogue of banding in a slow strain
rate sweep along the steady state flow curve.
Fig.~\ref{fig:pinPointMon} shows results for the nRP model with
parameters such that the underlying constitutive curve is monotonic.
Here steady state banding is absent in the limit $\omega\to 0$. In
both Fig.~\ref{fig:pinPointNonMon} and~\ref{fig:pinPointMon}, however,
significant banding is observed at high frequencies for a strain
amplitude $\gamma_0>1 $: this is the elastic banding associated with
the stress overshoot in each half cycle, described in detail above for
the $(\gdot_0,\omega)$ values denoted by $X_H$ in
Fig.~\ref{fig:sketch}.

It is important to emphasise, therefore, that even a fluid with a
purely monotonic constitutive curve, which does not shear band in
steady flow, is still nonetheless capable of showing strong shear
banding in a time-dependent protocol of high enough frequency
(Fig.~\ref{fig:pinPointMon}). Also important to note is that for a
fluid with a non-monotonic constitutive curve the region of steady
state `viscous' banding at low frequencies crosses over smoothly to
that of `elastic' banding as the frequency increases
(Fig.~\ref{fig:pinPointNonMon}). 

Corresponding to the degree of banding in the shear rate
$\delta\gdot$, as plotted in Figs.~\ref{fig:pinPointNonMon}
and~\ref{fig:pinPointMon}, is an equivalent degree of banding $G\delta
W_{xy}=-\eta\delta\gdot$ (to within small corrections of order the
cell curvature, $q$) in the shear component of the polymeric
conformation tensor. This follows trivially by imposing force balance
at zero Reynolds number. Counterpart maps for the degree of banding in
the component $\delta W_{yy}$ of the polymeric conformation tensor can
also be built up. These reveal closely similar regions of banding to
those shown in Figs.~\ref{fig:pinPointNonMon}
and~\ref{fig:pinPointMon}. (Data not shown.) Experimentally,
heterogeneities in the polymeric conformation tensor can be accessed
by flow birefringence.

\begin{figure}[tbp]
\includegraphics[width=8cm]{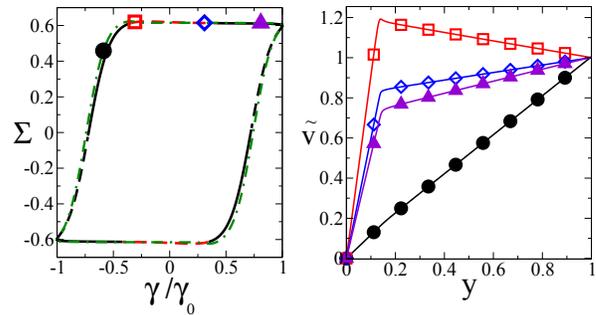}
\caption{As in Fig.~\ref{fig:portraitNonMon} but for the nRP model with
  a CCR parameter $\beta=1.0$ for which the underlying homogeneous
  constitutive curve is monotonic, and for a LAOStrain with strain
  rate amplitude $\gamma_{0}=56.2$ and frequency
  $\omega=10.0$. Number of numerical grid points $J=512$.}
\label{fig:portraitMon}
\end{figure}

As noted above, to build up such comprehensive roadmaps as in
Figs.~\ref{fig:pinPointNonMon} and~\ref{fig:pinPointMon} in a
computationally efficient way, we omitted all nonlinear effects and
integrated instead the linearised equations of Sec.~\ref{sec:lsa}.
These are only strictly valid in any regime where the amplitude of the
heterogeneity remains small. In omitting nonlinear effects, they tend
to overestimate the degree of banding in any regime of sustained
positive eigenvalue, in predicting the heterogeneity to grow
exponentially without bound, whereas in practice it would be cutoff by
nonlinear effects. We now remedy this shortcoming by exploring the
model's full nonlinear spatiotemporal dynamics. To do so within
feasible computational run times, we focus on a restricted grid of
values in the $\gdot_0,\omega$ plane, marked by crosses in
Figs.~\ref{fig:pinPointNonMon} and~\ref{fig:pinPointMon}.

The results are shown in Fig.~\ref{fig:PipkinNonMon} for the nRP model
with model parameters for which the underlying constitutive curve is
non-monotonic.  At low frequencies the results tend towards the
limiting fluid-like behaviour discussed above, in which the the stress
slowly tracks up and down the steady state flow curve $\Sigma(\gdot)$.
(Progression towards this limit can be seen by following the top row
of panels in Fig.~\ref{fig:PipkinNonMon}b) to the left.)  Viscous
banding is seen for sufficiently high strain rate amplitudes $\gdot_0$
due to the negatively sloping underlying homogeneous constitutive
curve.  At high frequencies the response tends instead towards the
limiting elastic-like behaviour discussed above. For large enough
strain amplitudes the stress then shows an open cycle $\Sigma(\gamma)$
as a function of strain, with an overshoot in each half cycle that
triggers the formation of `elastic' banding.  (Progression towards
this limit is seen by following the top row of panels in
Fig.~\ref{fig:PipkinNonMon}a to the right.)

Overshoots in the elastic Lissajous-Bowditch curve of stress as a
function of strain have been identified in earlier
work~\cite{Ewoldt2009loops} as leading to self-intersection of the
corresponding viscous Lissajous-Bowditch curve of stress as a function
of strain-rate. Such an effect is clearly seen here in the Rolie-poly
model: see for example the runs highlighted by the thicker boxes in
Fig.~\ref{fig:PipkinNonMon} and in Fig.~\ref{fig:PipkinMon}.

\begin{figure}[tbp]
\includegraphics[width=8.5cm]{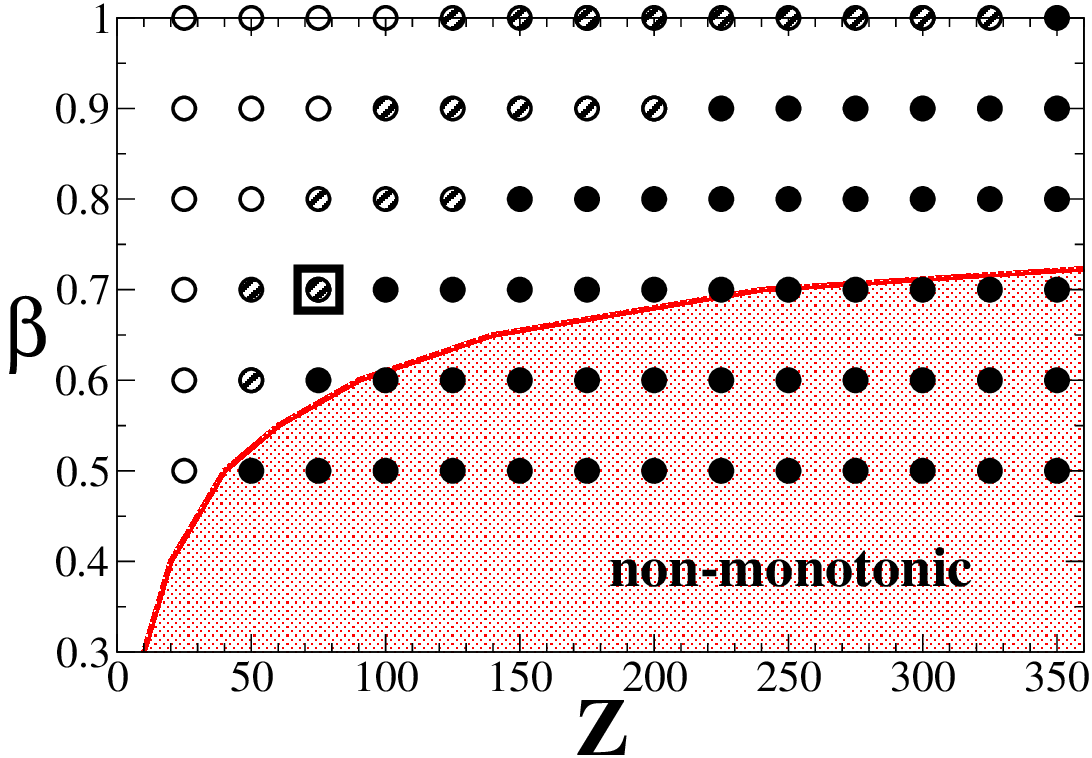}
\caption{Effect of CCR parameter $\beta$ and entanglement number $Z$
  (and so of chain stretch relaxation time $\taur=\taud/3Z$) on shear
  banding in LAOStrain.  (Recall that the non-stretching version of
  the model has $\taur\to 0$ and so $Z\to\infty$.)  Empty circles: no
  observable banding.  Hatched circles: observable banding, typically
  $\Delta_{\gdot}/\gdot_0 \approx 10\%-100\%$. Filled circles: significant
  banding $\Delta_{\gdot}/\gdot_0 \ge 100\%$.  For hatched and filled
  symbols we used the criterion that banding of the typical magnitude
  stated is apparent in a region spanning at least half a decade by
  half a decade in the plane of $\gdot_0,\omega$, by examining maps as
  in Fig.~\ref{fig:stretchPinPoint} in by eye. The square shows the
  parameter values explored in detail in
  Fig.~\ref{fig:stretchPinPoint}.}
\label{fig:stretchMaster}
\end{figure}

For intermediate frequencies the stress is a complicated function of
both strain rate and also, separately, the strain. The three
dimensional curve $(\Sigma,\gdot,\gamma)$ is then best shown as two
separate projections: first in the elastic representation of the
$\Sigma,\gamma$ plane (Fig.~\ref{fig:PipkinNonMon}a), and second, in
the viscous representation of the $(\Sigma,\gdot)$ plane
(Fig.~\ref{fig:PipkinNonMon}b).  Collections of these
Lissajous-Bowditch curves on a grid of $(\gdot_0,\omega)$ values as in
Fig.~\ref{fig:PipkinNonMon} are called Pipkin diagrams.

\begin{figure}[tbp]
   \includegraphics[width=10.0cm]{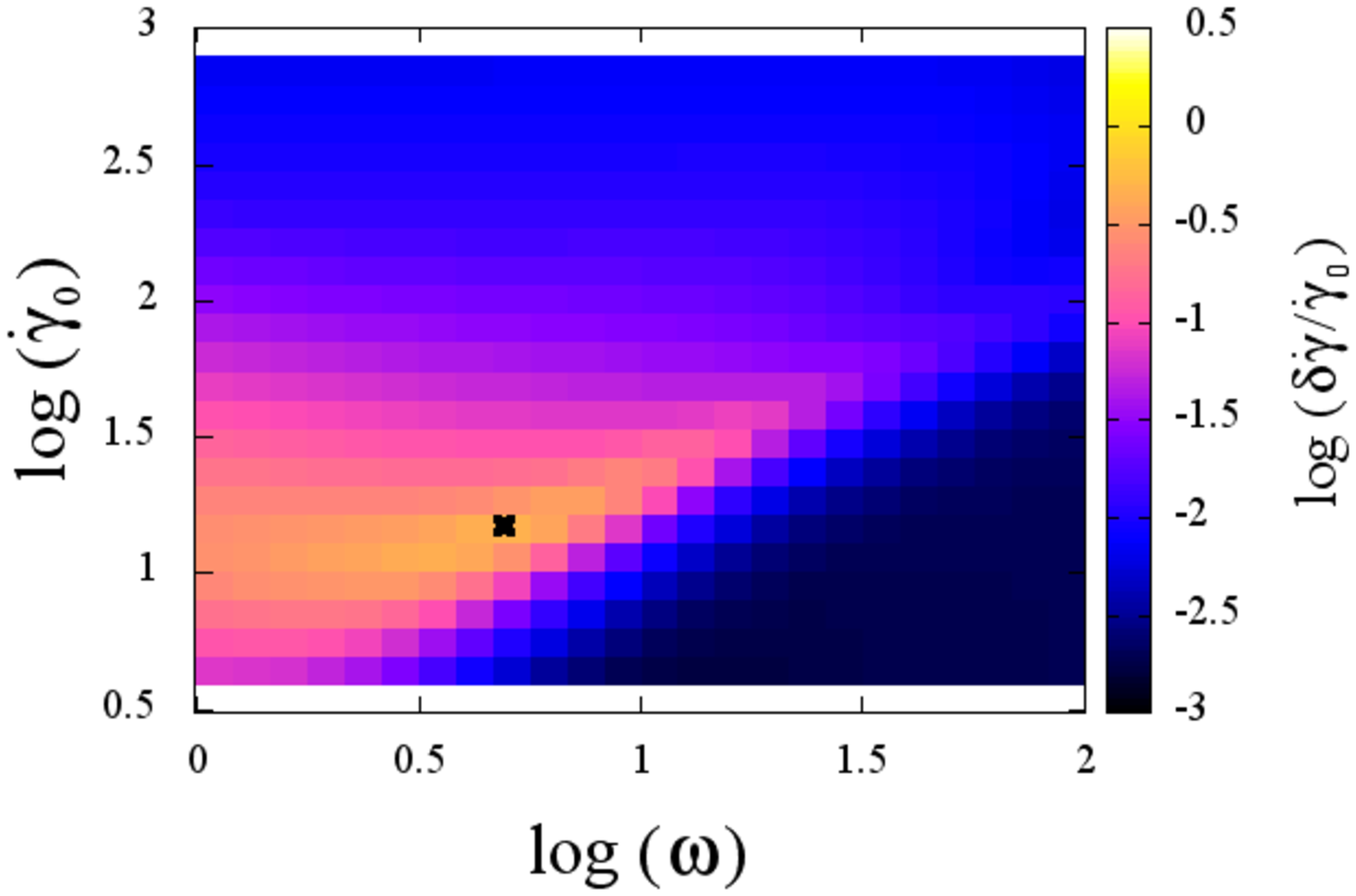}
\caption{Colour map of the normalised degree of shear banding for the
  sRP model with a monotonic constitutive curve. Each point in this
  $\gdot_0,\omega$ plane corresponds to a particular LAOStrain run
  with strain rate amplitude $\gdot_0$ and frequency $\omega$.  For
  computational efficiency, these calculations are performed by
  integrating the linearised equations in Sec.~\ref{sec:lsa}. Reported
  is the maximum degree of banding at any point in the cycle, after
  many cycles.  Model parameters: $\beta=0.7$, $Z=75$ (and so
  $\taur=0.0044$), $\eta=10^{-5}$.  Cell curvature $q=2\times
  10^{-3}$. Note the different colour scale from Figs.~\ref{fig:nonmon} and~\ref{fig:mon}. The model's full nonlinear dynamics for the ($\gdot_{0}, \omega$) value marked by the cross are explored in Fig.~\ref{fig:stretchPortrait}.}
\label{fig:stretchPinPoint}
\end{figure}

\begin{figure}[tbp]
\includegraphics[width=9cm]{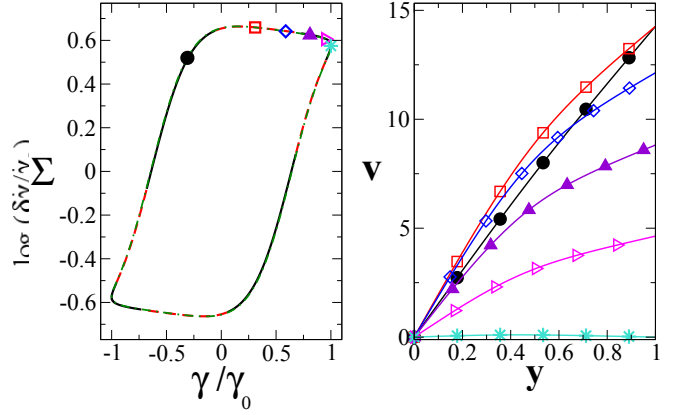}
\caption{sRP model with a monotonic constitutive curve in LAOStrain of
  strain rate amplitude $\gdot_{0}=20.0$ and frequency $\omega=8.0$.
  Model parameters $\beta=0.7, Z=75, \eta=10^{-5}$.  Cell curvature
  $q=2\times 10^{-3}$. Number of numerical grid points $J=512$. {\bf
    Left:} stress response in the elastic representation. Solid black
  and red-dashed line: calculation in which the flow is constrained to
  be homogeneous.  Red-dashed region indicates a positive eigenvalue
  showing instability to the onset of shear banding. Green dot-dashed
  line: stress response in a full nonlinear simulation that allows
  banding (almost indistinguishable from homogeneous signal in this
  case.) {\bf Right:} Velocity profiles corresponding to stages in the
  cycle indicated by matching symbols in left panel.}
\label{fig:stretchPortrait}
\end{figure}

For the particular LAOStrain run highlighted by a thicker box in
Fig.~\ref{fig:PipkinNonMon}, a detailed portrait of the system's
dynamics is shown in Fig.~\ref{fig:portraitNonMon}. Here the stress is
shown in the elastic representation, as a function of strain $\gamma$
(left hand panel). Two curves are shown here. The first shows the
stress signal in a calculation in which the flow is artificially
constrained to remain homogeneous. A linear instability analysis for
the dynamics of small heterogeneous perturbations about this
time-evolving homogeneous base state then reveals instability towards
the formation of shear bands (a positive eigenvalue) in the portion of
that curve shown as a red dashed line. A full nonlinear simulation
then reveals the formation of shear bands, and leads to a stress
signal (green dot-dashed line) that deviates from that of the
homogeneously constrained calculation, in particular in having a much
more precipitous stress drop due to the formation of bands.

The associated velocity profiles at four points round the part of the
cycle with increasing strain are shown in the right hand panel. Before
the stress overshoot, no banding is apparent (black circles). The
overshoot then triggers strong shear banding (red squares), with most
of the shear concentrated in a thin band at the left hand edge of the
cell. Interestingly, the shear in the right hand part of the cell is
in the opposite sense to the overall applied shear. This is consistent
with the fact that the stress is a decreasing function of strain in
this regime: the material is being unloaded, and an elastic-like
material being unloaded will shear backwards. As the overall applied
strain increases towards the end of the window of instability, the
flow heterogeneity gradually decays away. This process repeats in each
half cycle (with the obvious sign reversals in the part of the cycle
in which the strain is decreasing).

The corresponding Pipkin diagram for a fluid with a monotonic
constitutive curve (Fig.~\ref{fig:PipkinMon}) likewise confirms its
counterpart linear diagram in Fig.~\ref{fig:pinPointMon}. Here
`viscous' banding is absent at low frequencies, because the fluid is
not capable of steady state banding. Crucially, however, a strong
effect of elastic banding is still seen at high frequencies.  A
detailed portrait of the system's dynamics in this elastic regime, for
the strain rate amplitude and frequency marked by the thicker box in
Fig.~\ref{fig:PipkinMon}, is shown in Fig.~\ref{fig:portraitMon}. As
can be seen, it shows similar features to those just described in
Fig.~\ref{fig:portraitNonMon}. We emphasise, then, that even polymeric
fluids that do not band under conditions of steady shear are still
capable of showing strong banding in a time-dependent protocol at high
enough frequency. This important prediction is consistent with the
early insight of Adams and Olmsted in Ref.~\cite{Adams2009}.

So far we have presented results for the nRP model, which assumes an
infinitely fast rate of chain stretch relaxation compared to the rate
of reptation, such that the ratio $\taur/\taud\to 0$. This corresponds
to assuming that the polymer chains are very highly entangled, with a
number of entanglements per chain $Z=\taud/3\taur\to\infty$. We now
consider the robustness of these results to reduced entanglement
numbers, and accordingly increased chain stretch relaxation time
$\taur$ (in units of $\taud$). 

The results are summarised in Fig.~\ref{fig:stretchMaster}, which
shows the regions of the plane of the CCR parameter $\beta$ and
entanglement number $Z$ in which significant banding (filled circles),
observable banding (hatched circles), and no banding (open circles)
occur.  (Recall that results presented for the nRP model above pertain
to the values $\beta=0.4,1.0$ in the limit $Z\to\infty$.) As can be
seen, by reducing the number of entanglements per chain the effect of
shear banding is reduced and eventually eliminated.  However it is
important to note that, depending on the value of $\beta$, significant
banding is still observed for experimentally commonly used
entanglement numbers, typically in the range $20-100$. Furthermore, significant banding is seen in a large region of the ($\beta, Z$) plane for which the material's underlying constitutive curve is monotonic. As discussed
above, there is no current consensus as to the value of the CCR
parameter $\beta$ in the range $0<\beta<1$. Using the routemap
provided in Fig.~\ref{fig:stretchMaster}, a study of shear banding in
LAOStrain experiments could provide one way to obtain a more accurate
estimate of the value of this parameter.

For the pairing of $\beta$ and $Z$ values marked by the square in
Fig.~\ref{fig:stretchMaster}, we show in
Fig.~\ref{fig:stretchPinPoint} a colour map of the degree of banding
expected in LAOStrain in the space of strain rate amplitude $\gdot_0$
and frequency $\omega$. This figure, which is for the sRP model,
  is the counterpart of the earlier Figs.~\ref{fig:pinPointNonMon}
  and~\ref{fig:pinPointMon} discussed above for the nRP model, with
the degree of banding calculated for computational
efficiency within the assumption of linearised dynamics. Recall that
each pinpoint in this plane corresponds to a single LAOStrain run with
applied strain rate $\gdot(t)=\gdot_0\cos(\omega t)$.

Consistent with the underlying constitutive curve being monotonic for
these parameters, `viscous' banding is absence in the limit of low
frequencies $\omega\to 0$.  However significant banding is still
observed for runs with strain rate amplitudes $O(10-100)$ and
frequencies $O(1-10)$. This is the counterpart of the `elastic'
banding reported above in the nRP model, though the effect of finite
chain stretch in the sRP model is to moderate the degree of banding. A
detailed portrait of the model's nonlinear banding dynamics at a
strain rate amplitude $\gdot_0$ and frequency $\omega$ marked by the
cross in Fig.~\ref{fig:stretchPinPoint} is shown in
Fig.~\ref{fig:stretchPortrait}. Significant shear banding associated with the stress overshoot is apparent in each half cycle.

\section{Large amplitude oscillatory stress}
\label{sec:LAOStress}

We now consider the time-dependent stress-imposed oscillatory protocol
of LAOStress. Here the sample is subject for times $t>0$ to a stress
of the form
\beqn
\Sigma(t)=\Sigma_0\sin(\omega t),
\eeqn
characterised by the frequency $\omega$ and stress amplitude
$\Sigma_0$. As for the case of LAOStrain above, all the results
presented below are in the long-time regime, once many ($N=20$) cycles
have been executed and the response of the system has settled to be
time-translationally invariant from cycle to cycle, $t\to
t+2\pi/\omega$.

In Sec.~\ref{sec:recapCreep} we reviewed existing work demonstrating
the tendency to form shear bands in a {\rm step} stress experiment.
Here an initially well relaxed sample is suddenly subject at time
$t=0$ to the switch-on of a shear stress of amplitude $\Sigma_0$,
which is held constant for all subsequent times.  The criterion for an
underlying base state of initially homogeneous creep response to
become linearly unstable to the formation of shear bands is then that
the time-differentiated creep response curve $\gdot(t)$
obeys~\cite{Moorcroft2013}:
\beqn
\label{eqn:critCreep}
\frac{\partial^2\gdot}{\partial t^2}/\frac{\partial\gdot}{\partial t}>0.
\eeqn
Therefore, shear banding is expected in any regime where the
time-differentiated creep curve simultaneously slopes up and curves
upwards; or instead simultaneously slopes down and curves downwards.

\begin{figure}[tbp]
\includegraphics[width=8.0cm]{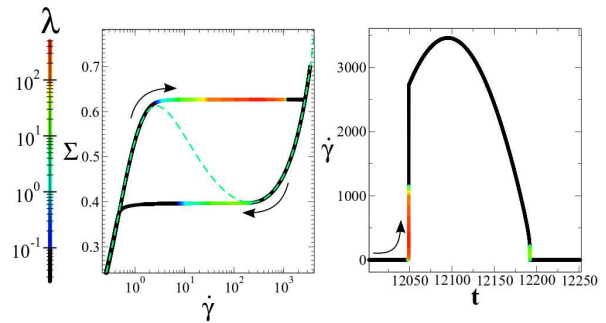}
\caption{LAOStress in the nRP model with a non-monotonic constitutive
  curve. Model parameters: $\beta=0.1$, $\eta=10^{-4}$. Frequency
  $\omega=0.01$ and stress amplitude $\Sigma_0=0.7$.  {\bf Left:}
  stress versus strain rate (shown on a log scale) in the positive
  stress part of the cycle. Colour scale shows eigenvalue, with
  negative values also shown as black. Green dashed line: underlying
  constitutive curve. {\bf Right:} corresponding stress versus time
  plot.}
\label{fig:eigenNonMon}
\end{figure}

\begin{figure}[tbp]
\includegraphics[width=8.0cm]{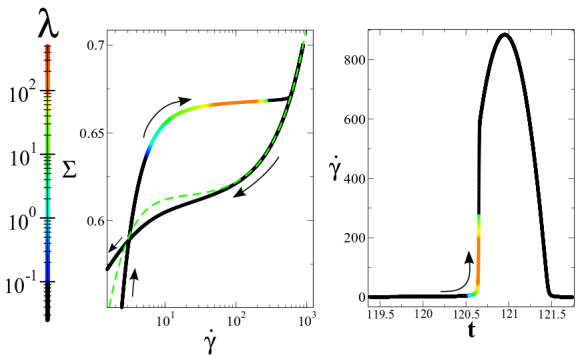}
\caption{As in Fig.~\ref{fig:eigenNonMon} but at a higher imposed
  frequency $\omega=1.0$ and for a value of the CCR $\beta=0.9$, for
  which the nRP model has a monotonic underlying constitutive curve.
  {\bf Right:} corresponding stress versus time plot.}
\label{fig:eigenMon}
\end{figure}

Having been derived within the assumption of an imposed stress that is
constant in time, criterion (\ref{eqn:critCreep}) would not {\it a
  priori} be expected to hold for the case of LAOStress. Nonetheless
it might reasonably be expected to apply, to good approximation, in any
regime of a LAOStress experiment where a separation of timescales
arises such that the shear rate $\gdot(t)$ evolves on a much shorter
timescale than the stress. In this case, from the viewpoint of the
strain rate signal, the stress appears constant in comparison and the
constant-stress criterion (\ref{eqn:critCreep}) is expected to hold.
Indeed, in what follows we shall show that many of our results for
LAOStress can be understood on the basis of this simple piece of
intuition.

We start in Fig.~\ref{fig:eigenNonMon} by considering the nRP model in
a parameter regime for which the underlying constitutive curve is
non-monotonic (see the dotted line in the left panel), such that shear
banding would be expected under conditions of a steadily applied shear
rate.  With the backdrop of this constitutive curve we consider a
LAOStress run at low frequency $\omega\to 0$. For definiteness we will
focus on the part of the cycle where the stress is positive, but
analogous remarks will apply to the other half of the cycle, with
appropriate changes of sign.

Consider first the regime in which the stress is slowly increased from
$0$ towards its maximum value $\Sigma_0$. In this part of the cycle,
at the low frequencies of interest here, we expect the system to
initially follow the high viscosity branch of the constitutive curve.
In any experiment for which the final stress $\Sigma_0$ exceeds the
local maximum in the constitutive curve, the system must at some stage
during this increasing-stress part of the cycle transit from the high
to low viscosity branch of the constitutive curve. This transition is
indeed seen in Fig.~\ref{fig:eigenNonMon}: it occurs via ``top
jumping'' from the stress maximum across to the low viscosity branch.
Conversely, on the downward part of the sweep as the stress decreases
from its maximum value $\Sigma_0$, the system initially follows the
low viscosity branch until it eventually jumps back to the high
viscosity branch. (We return in our concluding remarks to discuss the
possible effect of thermal nucleation events, which are not included
in these simulations, on these process of jumping between the two
branches of the constitutive curve.)

The corresponding signal of strain rate versus time during this slow
up-then-down stress oscillation is shown in the right panel of
Fig.~\ref{fig:eigenNonMon}. As can be seen, the regimes where the
shear rate transits between the two different branches of the
constitutive curve occur over relatively short time intervals. (The
duration of this process is informed by the short timescale $\eta/G$,
whereas the stress evolves on the much longer timescale $2\pi/\omega$.)
This separation of timescales renders the stress signal approximately
constant in comparison to the fast evolution of the strain rate.
Criterion (\ref{eqn:critCreep}) might therefore be expected to apply
in this regime of transition, at least to good approximation.

Furthermore, during the transition from the high to low viscosity
branch we see a regime in which the shear rate signal simultaneously
slopes up and curves up as a function of time: criterion
(\ref{eqn:critCreep}) not only applies but is also met, and we
therefore expect an instability to banding.  Plotting, by means of a
colourscale, the eigenvalue as defined in Sec.~\ref{sec:lsa}, we find
that it is indeed positive. Likewise, during the rapid transition back
from the low to high viscosity branch, we find a regime in which the
shear rate signal simultaneously slopes down and curves down. As seen
from the colourscale, the eigenvalue is also positive in this regime
(although more weakly than during the upward transition). In what
follows, we will confirm the expectation of shear band formation
during these times of positive eigenvalue by simulating the model's
full nonlinear spatiotemporal dynamics.

\begin{figure}[tbp]
\includegraphics[width=9.5cm]{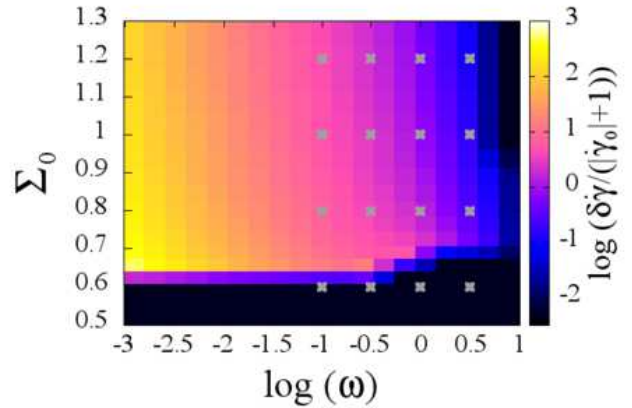}
\caption{Colour map of the normalised degree of shear banding for the
  nRP model with a non-monotonic constitutive curve. Each point in
  this $\Sigma_0,\omega$ plane corresponds to a particular LAOStress
  run with stress amplitude $\Sigma_0$ and frequency $\omega$.  For
  computational efficiency, these calculations are performed by
  integrating the linearised equations in Sec.~\ref{sec:lsa}. Reported
  is the maximum degree of banding that occurs at any point in the
  cycle, after many cycles.  Model parameters: $\beta=0.4$,
  $\eta=10^{-4}$.  Cell curvature $q=2\times 10^{-3}$. Crosses
  indicate the grid of values of $\Sigma_{0}$ and $\omega$ in
  Fig.~\ref{fig:PipkinNonMon2}.}
\label{fig:pinPointNonMon2}
\end{figure}

\begin{figure}[tbp]
\includegraphics[width=9.5cm]{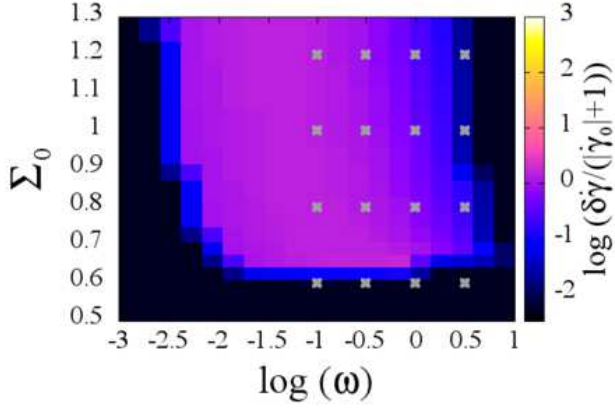}
\caption{ As in Fig.~\ref{fig:pinPointNonMon}, but with a CCR
  parameter $\beta=0.9$, for which the fluid has a monotonic
  underlying constitutive curve.  Crosses indicate the grid of values
  of $\Sigma_{0}$ and $\omega$ used in the Pipkin diagram of
  Fig.~\ref{fig:PipkinMon2}.}
\label{fig:pinPointMon2}
\end{figure}

These processes of rapid transition between different branches of the
constitutive curve are of course not expected in a LAOStress
experiment at low frequency for a fluid with a monotonic constitutive
curve. In this case, for a LAOStress run in the limit $\omega\to 0$
the system instead sweeps quasi-statically along the monotonic
constitutive curve, with no associated banding.  As in the case of
LAOStrain, however, it is crucial to realise that the absence of
banding in an experiment at zero frequency does not preclude the
possibility of banding in a time-dependent protocol at finite
frequency, even in a fluid with a monotonic constitutive curve.

\begin{figure}[tbp]
\includegraphics[width=8.5cm]{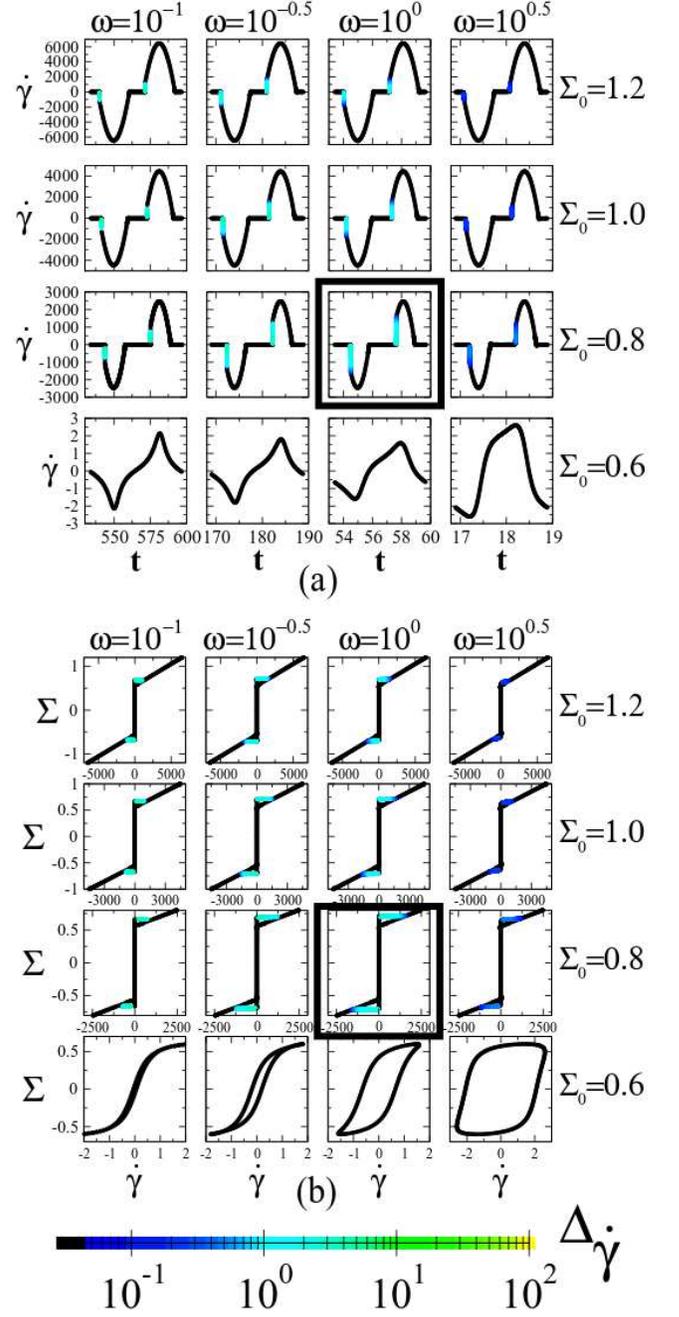}
\caption{Lissajous-Bowditch curves in LAOStress for the nRP model with
  a non-monotonic constitutive curve. Results are shown as shear-rate
  versus time in (a), and in the viscous representation of stress
  versus strain rate in (b). Columns of fixed frequency $\omega$ and
  rows of fixed strain-rate amplitude $\gdot_0$ are labeled at the top
  and right-hand side. Colourscale shows the time-dependent degree of
  shear banding.  Model parameters: $\beta=0.4, \eta=10^{-4}, l=0.02$.
  Cell curvature: $q=2\times 10^{-3}$. Number of numerical grid points $J=512$. A detailed portrait of the run
  outlined by the thicker box is shown in
  Fig.~\ref{fig:portraitNonMon2}.}
\label{fig:PipkinNonMon2}
\end{figure}

\begin{figure}[tp!]
\includegraphics[width=8.5cm]{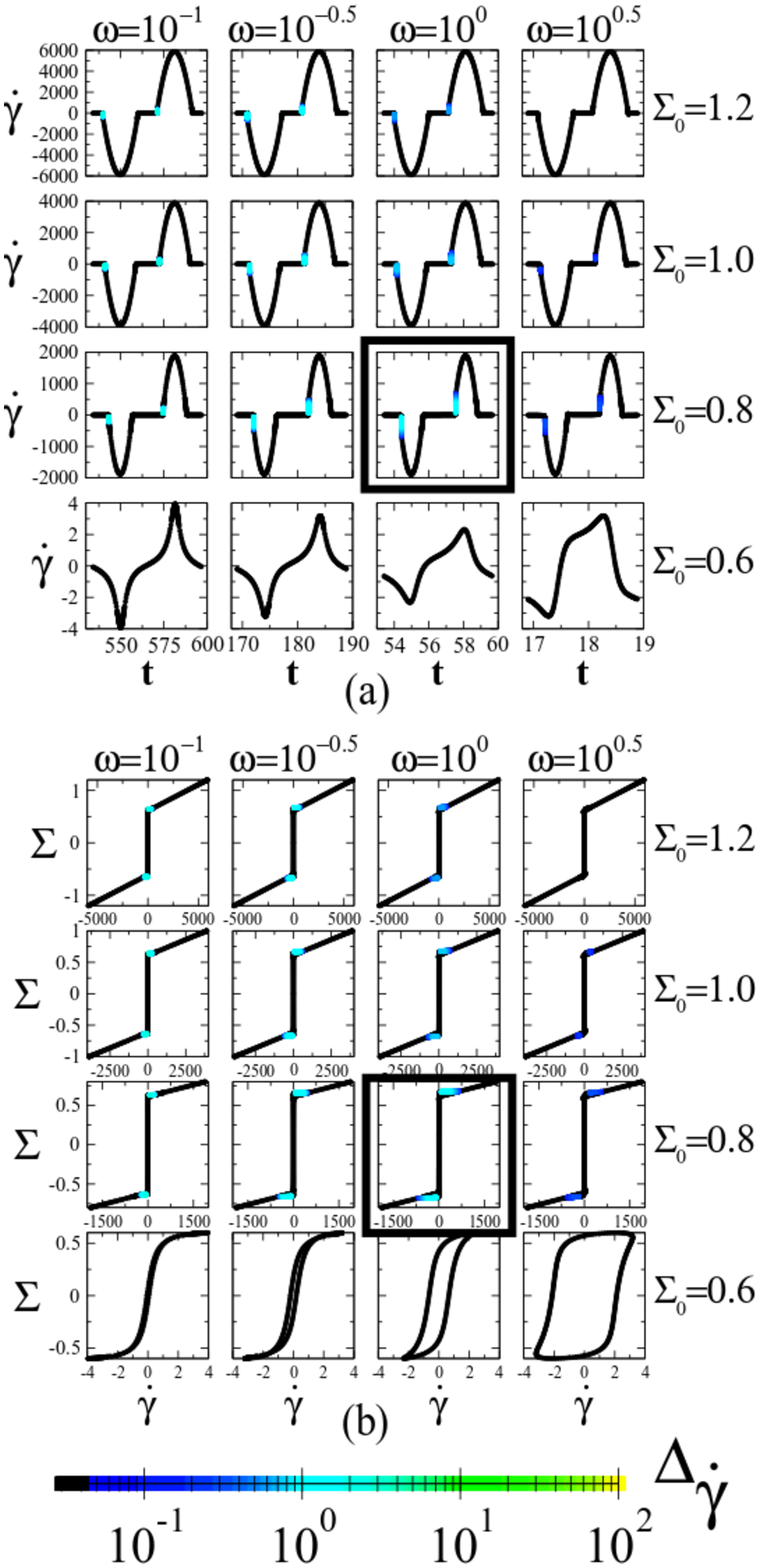}
\caption{As in Fig.~\ref{fig:PipkinNonMon2} but for a value of the CCR
  parameter $\beta=0.9$, for which the fluid's underlying constitutive
  curve is monotonic. Number of numerical grid points $J=512$. A detailed portrait of the run outlined by the thicker box is shown in Fig.~\ref{fig:portraitMon2}.}
\label{fig:PipkinMon2}
\end{figure}

With that in mind, we show in Fig.~\ref{fig:eigenMon} the counterpart
of Fig.~\ref{fig:eigenNonMon}, but now for the nRP model with a
monotonic constitutive curve subject to a LAOStress run at a finite
frequency $\omega=1$, of order the fluid's reciprocal stress
relaxation timescale.  The key to understanding the emergent dynamics
in this case is the existence in the underlying zero-frequency
constitutive curve (shown by a dotted line in the left panel) of a
region in which the stress is a relatively flat (though still
increasing) function of the strain rate. As the system transits this
region during the increasing stress part of a finite-frequency stress
cycle, we again observe a regime of quite sudden progression from low
to high strain rate. This is seen in the left to right transition in
the stress versus strain rate representation in the left panel of
Fig.~\ref{fig:eigenMon}, and (correspondingly) in the rapid increase
of strain rate versus time in the right panel.

During this regime of rapid transit we again have conditions in which
the strain rate evolves rapidly compared to the stress, such that the
constant-stress criterion (\ref{eqn:critCreep}) should apply to good
approximation. Furthermore, during the first part of the transition,
the strain rate signal simultaneously slopes and curves upward as a
function of time.  The eigenvalue is therefore positive, indicating
linear instability of an initially homogeneous base state to the
formation of shear bands. We will again confirm this prediction by
simulating the model's full nonlinear spatiotemporal dynamics below.

In the context of Figs.~\ref{fig:eigenNonMon} and~\ref{fig:eigenMon}
we have discussed the dynamics of the nRP model with a non-monotonic
and monotonic constitutive curve respectively, focusing in each case
on one particular value of the imposed frequency $\omega$ and stress
amplitude $\Sigma_0$. We now consider the full plane of
$(\Sigma_0,\omega)$ by showing in Figs.~\ref{fig:pinPointNonMon2}
and~\ref{fig:pinPointMon2} colour maps of the extent of banding across
it.  Recall that each point in this plane corresponds to a single LAOS
experiment with stress amplitude $\Sigma_0$ and frequency $\omega$.
To build up these maps we sweep over a grid of 20x20 values of
$\Sigma_0,\omega$ and execute at each point a LAOStress run,
integrating the model's linearised equations set out in
Sec.~\ref{sec:lsa}. We then represent the degree of banding
$\delta\gdot$, maximised over the cycle after many cycles, by the
colourscale.

\begin{figure}[tbp]
\includegraphics[width=9cm]{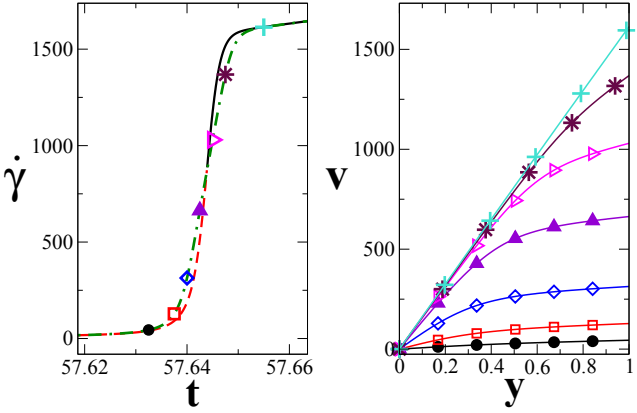}
\caption{ LAOStress in the nRP model with a non-monotonic constitutive
  curve. Stress amplitude $\Sigma_{0}=0.8$ and frequency $\omega=1.0$.
  Model parameters $\beta=0.4, \eta=10^{-4}, l=0.02$.  Cell curvature
  $q=2\times 10^{-3}$. Number of numerical grid points $J=512$. {\bf
    Left:} strain rate response as a function of time, focusing on the
  region in which the system transits from the high to low viscosity
  branch of the constitutive curve. Solid black and red-dashed line:
  calculation in which the flow is constrained to be homogeneous.
  Red-dashed region indicates a positive eigenvalue showing
  instability to the onset of shear banding. Green dot-dashed line:
  stress response in a full nonlinear simulation that allows banding.
  {\bf Right:} Velocity profiles corresponding to stages in the cycle
  indicated by matching symbols in the left panel.}
\label{fig:portraitNonMon2}
\end{figure}

Fig.~\ref{fig:pinPointNonMon2} shows results with model parameters for
which the underlying constitutive curve is non-monotonic. As expected,
for stress amplitudes $\Sigma_0$ exceeding the local maximum in the
underlying constitutive curve, significant banding is observed even in
the limit of low frequency $\omega\to 0$. This is associated with the
processes of jumping between the two different branches of the
constitutive curve discussed above.  

Fig.~\ref{fig:pinPointMon2} shows results for the nRP model with a
monotonic underlying constitutive curve. Here steady state banding is
absent in the limit $\omega\to 0$, as expected.  However, significant
banding is still nonetheless observed at frequencies of order the
reciprocal reptation time, for imposed stress amplitudes exceeding the
region of weak slope in the constitutive curve, consistent with our
discussion of Fig.~\ref{fig:eigenMon} above.

To obtain the comprehensive roadmaps of
Figs.~\ref{fig:pinPointNonMon2} and~\ref{fig:pinPointMon2} in a
computationally efficient way, we discarded any nonlinear effects and
integrated the linearised model equations set out in
Sec.~\ref{sec:lsa}. However, these linearised equations tend to
overestimate the degree of banding in any regime of sustained positive
eigenvalue.  Therefore in Figs.~\ref{fig:PipkinNonMon2}
and~\ref{fig:PipkinMon2} we now simulate the model's full nonlinear
spatiotemporal dynamics, restricting ourselves for computational
efficiency to grids of 4x4 values of $\Sigma_0,\omega$ as marked by
crosses in Figs.~\ref{fig:pinPointNonMon2} and~\ref{fig:pinPointMon2}.

Fig.~\ref{fig:PipkinNonMon2} pertains to the nRP model with model
parameters for which the underlying constitutive curve is
non-monotonic.  At low frequencies the results tend towards the
limiting behaviour discussed above, in which the stress slowly tracks
up and down the steady state flow curve $\Sigma(\gdot)$ in between
regimes of sudden transition between the two branches of the curve,
during which shear bands form. This is most pronounced in the case of
the jump between the high and low viscosity branches during the upward
sweep.  Banding on the downward sweep is only apparent in a relatively
more limited region of $\Sigma_0,\gdot$ space, consistent with the
transition of $\gdot_0$ being more modest in this part of the cycle
during which the stress decreases.

\begin{figure}[tbp]
\includegraphics[width=9cm]{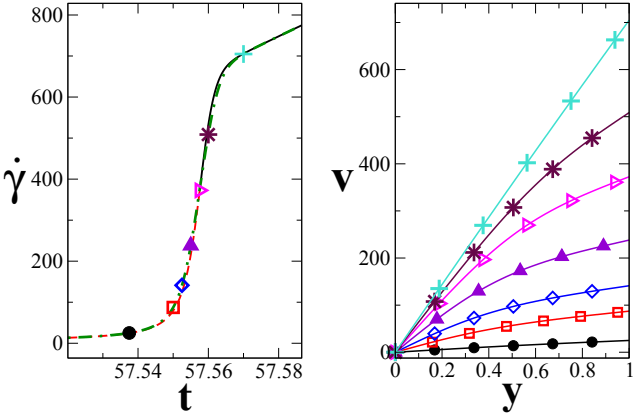}
\caption{As in Fig.~\ref{fig:portraitNonMon2} but for the nRP model
  with a CCR parameter $\beta=0.9$ for which the underlying
  homogeneous constitutive curve is monotonic. Number of numerical
  grid points $J=512$.}
\label{fig:portraitMon2}
\end{figure}

For the particular run highlighted by the thicker box in
Fig.~\ref{fig:PipkinNonMon2}, a detailed portrait of the system's
dynamics is shown in Fig.~\ref{fig:portraitNonMon2}. The left panel
shows the strain rate signal as a function of time, zoomed on the
region in which the strain rate makes its transit from the high to low
viscosity branch of the constitutive curve. The black and red-dashed
line show the results of a calculation in which the flow is
artificially constrained to remain homogeneous. The red-dashed region
indicates the regime in which the criterion (\ref{eqn:critCreep}) for
linear instability to the formation of shear bands is met, which also
corresponds to the regime in which the strain rate signal
simultaneously slopes up and curves upwards.

In a simulation that properly takes account of flow heterogeneity,
shear bands indeed develop during this regime where the criterion is
met, then decay again once the strain rate signal curves down and
stability is restored. This sequence can be seen in the velocity
profiles in the right hand panel. The stress signal associated with
this run that allows bands to form is shown by the green dot-dashed
line in the left panel, and is only barely distinguishable from that
of the run in which the flow is constrained to remain homogeneous.

Fig.~\ref{fig:PipkinMon2} pertains to the nRP model with model
parameters for which the underlying constitutive curve is monotonic,
with the grid of $(\Sigma_0,\omega)$ values that it explores marked by
crosses in Fig.~\ref{fig:pinPointMon2}. True top-jumping events are
absent here, and no shear banding arises in the limit of zero
frequency. As discussed above, however, a similar rapid transition
from low to high shear rate is seen in runs at a frequency $O(1)$, as
the stress transits the region of weak slope in the constitutive curve
during the increasing-stress part of the cycle.  Associated with this
transit is a pronounced tendency to form shear bands.

\begin{figure}[tbp]
\includegraphics[width=8.5cm]{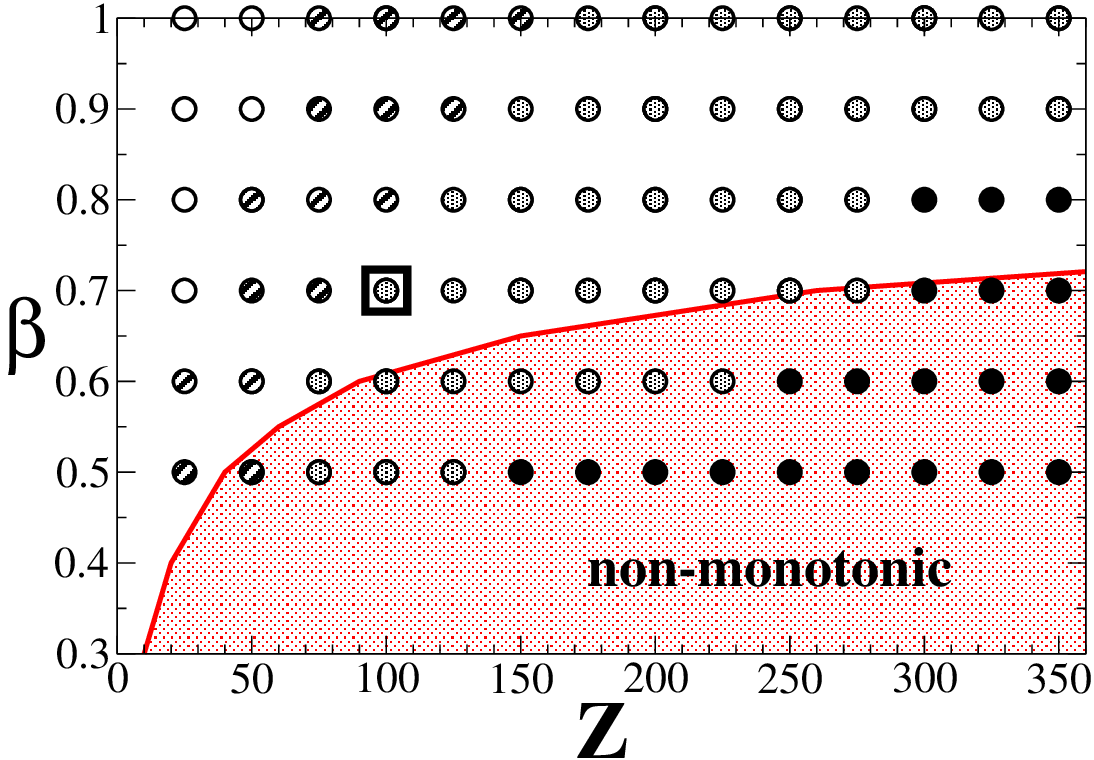}
\caption{Effect of CCR parameter $\beta$ and entanglement number $Z$
  (and so of chain stretch relaxation time $\taur=\taud/3Z$) on shear
  banding in LAOStress.  (Recall that the non-stretching version of
  the model has $\taur\to 0$ and so $Z\to\infty$.)  Empty circles: no
  observable banding.  Hatched circles: observable banding,
  $\Delta_{\gdot}/(1+|\gdot(t)|) = 10\%-31.6\%$. Dot-filled circles:
  significant banding, $\Delta_{\gdot}/(1+|\gdot(t)|) = 31.6\% - 100\%$.
  Filled circles: strong banding, $\Delta_{\gdot}/(1+|\gdot(t)|) > 100\%$.
  For the hatched, dot-filled and filled symbols we used the criterion
  that banding of the typical magnitude stated is apparent for any of
  $\omega=0.1,0.316$ or $1.0$, given a stress amplitude $\Sigma_0$
  exceeding the region of weak slope in the constitutive curve. The
  square shows the parameter values explored in detail in
  Fig.~\ref{fig:stretchPortrait1}. The solvent viscosity $\eta$ is $3.16\times10^{-5}$.}
\label{fig:stretchMaster2}
\end{figure}

This can be seen for the run highlighted by the thicker box in
Fig.~\ref{fig:PipkinMon2}, of which a detailed portrait is shown in
Fig.~\ref{fig:portraitMon2}. This shows very similar features to its
counterpart for a non-monotonic underlying constitutive curve. In
particular, the regime of simultaneous upward slope and upward
curvature in the strain rate signal as the stress transits the region
of weak positive constitutive slope triggers pronounced shear banding.

These results illustrate again the crucial point: that shear bands can
form in a protocol with sufficiently strong time-dependence, even in a
fluid for which the underlying constitutive curve is monotonic such
that banding is forbidden in steady state flows.

So far, we have restricted our discussion of LAOStress to the nRP
model, for which the stretch relaxation time $\taur$ is set to zero
upfront so that any chain stretch relaxes to zero instantaneously,
however strong the applied flow. The results of these calculations are
expected to apply, to good approximation, to experiments performed in
flow regimes where chain stretch remains small. This typically imposes
the restriction $\gdot\taur\ll 1$. We now turn to the sRP model to
consider the effects of finite chain stretch in experiments where this
restriction is not met.

\begin{figure}[tbp]
\includegraphics[width=8.5cm]{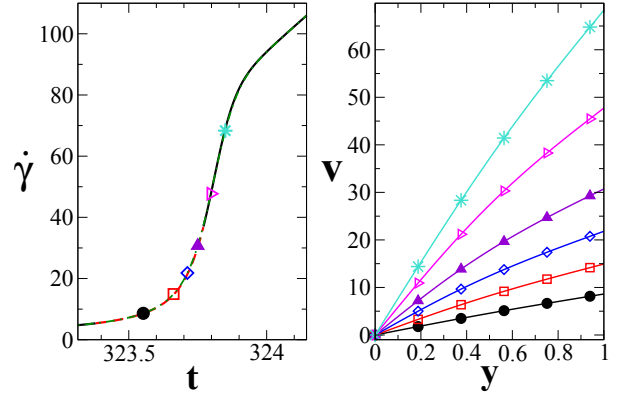}
\caption{sRP model with a monotonic constitutive curve in LAOStress of
  stress amplitude $\Sigma_{0}=0.8$ and frequency $\omega=0.1$.
  Model parameters $\beta=0.7, Z=100, \eta=3.16\times10^{-5}$.  Cell curvature
  $q=2\times 10^{-3}$. Number of numerical grid points $J=512$. {\bf
    Left:} strain rate signal versus time. Solid black
  and red-dashed line: calculation in which the flow is constrained to
  be homogeneous.  Red-dashed region indicates a positive eigenvalue
  showing instability to the onset of shear banding. Green dot-dashed
  line: stress response in a full nonlinear simulation that allows
  banding (indistinguishable from homogeneous signal in this
  case.) {\bf Right:} Velocity profiles corresponding to stages in the
  cycle indicated by matching symbols in left panel.}
\label{fig:stretchPortrait1}
\end{figure}

Fig.~\ref{fig:stretchMaster2} shows the regions of the plane of the CCR
parameter $\beta$ and entanglement number $Z$ in which banding can be
expected even with chain stretch.  As for the case of LAOStrain above
we note that, depending on the value of $\beta$, significant banding
is still observed for experimentally commonly used entanglement
numbers, typically in the range $20-100$. Furthermore, observable
banding is clearly evident over a large region of this plane in which
the underlying constitutive curve is monotonic, precluding steady
state banding. Again, we hope that this figure might act as a roadmap
to inform the discussion concerning the value of the CCR parameter
$\beta$.

For the pairing of $\beta$ and $Z$ values marked by the square in
Fig.~\ref{fig:stretchMaster2}, we show in
Fig.~\ref{fig:stretchPortrait1} a detailed portrait of the model's nonlinear dynamics at a stress amplitude $\Sigma_0$ and frequency $\omega$ for which observable banding occurs.

\section{Conclusions}
\label{sec:conclusions}

We have studied theoretically the formation of shear bands in large
amplitude oscillatory shear (LAOS) in the Rolie-poly model of polymers and wormlike micellar surfactants, with the particular aims of
identifying the regimes of parameter space in which shear banding is
significant, and the mechanisms that trigger its onset.

At low frequencies, the protocol of LAOStrain effectively corresponds
to a repeating series of quasi-static sweeps up and down the steady
state flow curve.  Here, as expected, we see shear banding in those
regimes of parameter space for which the fluid's underlying
constitutive curve is non-monotonic, for strain rate amplitudes large
enough to enter the banding regime in which the stress is a
characteristically flat function of strain rate.

In LAOStrain at higher frequencies we report banding not only in the
case of a non-monotonic constitutive curve, but also over a large
region of parameter space for which the constitutive curve is
monotonic and so precludes steady state banding. We emphasise that
this is an intrinsically time-dependent banding phenomenon that would
be absent under steady state conditions, and we interpret it as the
counterpart of the `elastic' banding predicted recently in the context
of shear startup experiments at high strain rates~\cite{Moorcroft2013}.

In LAOStress we report shear banding in those regimes of parameter
space for which the underlying constitutive curve is either negatively
or weakly positively sloping. In this case, the bands form during the
process of yielding associated with the dramatic increase in shear
rate that arises during that part of the cycle in which the stress magnitude
transits the regime of weak constitutive slope in an upward direction.
Although the banding that we observe here is dramatically apparent
during those yielding events, these events are nonetheless confined to
a relatively small part of the stress cycle as a whole and would
therefore need careful focus to be resolved experimentally. (A
possible related protocol, more focused on the banding regime,
could be to perform a shifted stress oscillation
$\Sigma(t)=\Sigma_{\rm plat} + \Delta\Sigma\sin(\omega t)$ where
$\Sigma_{\rm plat}$ is a characteristic stress value in the region of
weak slope in the constitutive curve and $\Delta\Sigma$ smaller in
comparison.) 

The dramatic increase in strain rate associated with transiting to the
high shear branch in LAOStress is likely to present practical
experimental difficulties in open flow cells such as Couette or
cone-and-plate. To circumvent this, flow in a closed microfluidic channel provides an attractive alternative to those macroscopic
geometries in seeking to access this effect experimentally. 

In each case we have demonstrated that the onset of shear banding can,
for the most part, be understood on the basis of previously derived
criteria for banding in simpler time-dependent
protocols~\cite{Moorcroft2013}. In particular, the trigger for
banding in LAOStrain at low frequencies is that of a negatively
sloping stress versus strain rate, which has long been recognised as
the criterion for banding under conditions of a steady applied shear
flow. The trigger in LAOStrain at high frequencies is instead that of
an overshoot in the signal of stress as a function of strain, in close
analogy to the criterion for banding onset during a fast shear startup
run. The trigger for banding in LAOStress is that of a regime of
simultaneous upward slope and upward curvature in the
time-differentiated creep response curve of strain rate as a function
of time. This again is a close counterpart to the criterion for
banding following the imposition of a step stress.

For both LAOStrain and LAOStress we have provided a map of shear
banding intensity in the space of entanglement number $Z$ and CCR
parameter $\beta$. We hope that this will provide a helpful roadmap
experimentalists, and might even help to pin down the value of the CCR
parameter, for which no consensus currently exists. 

We have also commented that the value of the Newtonian viscosity
$\eta$ is typically much smaller than the zero shear viscosity $G\tau$
of the viscoelastic contribution, giving $\eta\ll 1$ in our units.
Experimental literature suggests a range $\eta=10^{-7}$ to $10^{-3}$.
We have typically used $\eta=10^{-5}$ or
$\eta=10^{-4}$ in our numerics, and noted that the degree of banding
tends to increase with decreasing $\eta$. We also noted that the
timescale to transit from the high to low viscosity branch during
yielding in each half cycle in LAOStress decreases with decreasing
$\eta/G$. In view of these facts, a study of time-dependent banding in
fluids with smaller values of $\eta$ than those used here might
warrant the inclusion of inertia, because the small timescale for the
propagation of momentum might exceed the short timescale $\eta/G$ in
those cases.

In all our numerical studies the initial seed triggering the formation
of shear bands was taken to be the weak curvature that is present in
commonly used experimental flow cells. In order to demonstrate the
principle that the banding we report requires only a minimal seed,
rather than being an artefact of strong cell curvature, all our runs
have assumed a curvature that is much smaller than that of most flow
cells in practice. We also neglected stochastic noise altogether in
all the results presented here. (We have nonetheless also performed
runs with small stochastic noise instead of cell curvature and find
qualitatively all the same effects.)

However, one obvious shortcoming to this approach of taking only a very
small initial seed is that it tends to suppress the nucleation events
that are, in a real experimental situation, likely to trigger banding
even before the regime of true linear
instability~\cite{Grand1997}, particularly in low frequency
runs.  It would therefore be interesting in future work to study the
effect of a finite temperature with particular regard to the
nucleation kinetics to which it would give rise.

The calculations performed in this work all assumed from the outset
that spatial structure develops only in the flow gradient direction,
imposing upfront translational invariance in the flow and vorticity
directions. We defer to future work a study of whether, besides the
basic shear banding instabilities predicted here, secondary
instabilities~\cite{Fardin2014} of the interface between the
bands~\cite{Nghe2010,Fielding2005}
or of the high shear band itself~\cite{Fielding2010} will have
time to form in any given regime of amplitude and frequency space.

We have ignored throughout the effects of spatial variations in
  the concentration field. However, it is well known that in a
  viscoelastic solution heterogeneities in the flow field, and in
  particular in the normal stresses, can couple to the dynamics of
  concentration fluctuations via a positive feedback mechanism that
  enhances the tendency to form shear
  bands~\cite{Milner1993,Schmitt1995,Fielding2003a,Fielding2003b,Fielding2003c}.
  In the calculations performed here in LAOS we have observed
  significant differences in the viscoelastic normal stresses between
  the bands (approaching $50-70\%$ of the cycle-averged value of the
  same quantity, at least in the calculations without chain stretch).
  It would therefore clearly be interesting in future work to consider
  the effects of concentration coupling on the phenomena reported
  here.

Throughout we have ignored the possibility of edge fracture,
  because the one-dimensional calculations performed here lack any
  free surfaces and are unable to address it. It would clearly be
  interesting in future work to address the effects of edge fracture
  with regards the phenomena considered here~\cite{Skorski2011,
    Li2013, Li2015}.

All the calculations performed here have adopted what is
  essentially a single-mode approach, taking account of just one
  reptation relaxation timescale $\taud$ and one stretch relaxation
  timescale $\taur$. It would be interesting in future work to
  consider the effect of multiple relaxation timescales, which is
  likely to be an important feature of the dynamics of unbreakable
  polymers. (In wormlike micelles, in contrast, chain breakage and
  recombination narrows the relaxation spectrum significantly such
  that the single-mode approach adopted here is already likely to
  provide a reasonably full picture.)

We hope that this work will stimulate further experimental studies of
shear banding in time-dependent flows of complex fluids, with a
particular focus on the concept that banding is likely to arise rather
generically during yielding-like events (following a stress overshoot
in strain controlled protocols, or during a sudden increase in strain
rate in stress controlled protocols) even in fluids with a monotonic
constitutive curve that precludes steady state banding in a
continuously applied shear. In polymers this could form part of the
lively ongoing debate concerning the presence or otherwise of shear
banding in those materials. In wormlike micelles it would be
interesting to see a study of LAOS across the full phase diagram (as
set out, for example, in Ref.~\cite{Berret1997}), from
samples that band in steady state to those above the dynamical
critical point, which don't.

{\it Acknowledgements} The authors thank Alexei Likhtman, Elliot
Marsden, Peter Olmsted, Rangarajan Radhakrishnan, Daniel Read and
Dimitris Vlassopoulos for interesting discussions. The research
leading to these results has received funding from the European
Research Council under the European Union's Seventh Framework
Programme (FP/2007-2013), ERC grant agreement number 279365.


\end{document}